\documentclass[twocolumn, times, trackchanges]{aastex63}

\usepackage{gensymb}
\usepackage{booktabs}
\usepackage{rotating}

 \usepackage{float}
\usepackage{graphicx}
\usepackage{xkeyval}
\usepackage{ifpdf}
\usepackage{collectbox}

\usepackage[yyyymmdd,hhmmss]{datetime}

\received{08/03/2020}
\revised{10/02/2020}
\accepted{10/07/2020}
\submitjournal{ApJ}

\shorttitle{Nanoflare Brightenings in Flare Stars}
\shortauthors{Dillon et al.}

  \turnoffeditone   

\begin{document}

   \title{Statistical Signatures of Nanoflare Activity. II. \\ A Nanoflare Explanation for Periodic Brightenings in Flare Stars observed by NGTS}

\correspondingauthor{C.~J. Dillon}
\email{cdillon06@qub.ac.uk}

\author[0000-0003-2709-7693]{C.~J. Dillon}
\affiliation{Astrophysics Research Centre, School of Mathematics and Physics, Queen’s University Belfast, Belfast, BT7 1NN, UK}

\author[0000-0002-9155-8039]{D.~B. Jess}
\affiliation{Astrophysics Research Centre, School of Mathematics and Physics, Queen’s University Belfast, Belfast, BT7 1NN, UK}
\affiliation{Department of Physics and Astronomy, California State University Northridge, Northridge, CA 91330, USA}

\author[0000-0002-7725-6296]{M.~Mathioudakis}
\affiliation{Astrophysics Research Centre, School of Mathematics and Physics, Queen’s University Belfast, Belfast, BT7 1NN, UK}

\author[0000-0002-9718-3266]{C.~A. Watson}
\affiliation{Astrophysics Research Centre, School of Mathematics and Physics, Queen’s University Belfast, Belfast, BT7 1NN, UK}

\author[0000-0003-0711-7992]{J.~A.~G. Jackman}
\affiliation{Department of Physics, University of Warwick, Gibbet Hill Road, Coventry CV4 7AL, UK}

\author[0000-0003-1452-2240]{P.~J. Wheatley}
\affiliation{Department of Physics, University of Warwick, Gibbet Hill Road, Coventry CV4 7AL, UK}

\author{M.~R.Goad}
\affiliation{School of Physics and Astronomy, University of Leicester, University Road, Leicester, LE1 7RH, UK}

\author[0000-0003-2478-0120]{S.~L.Casewell}
\affiliation{School of Physics and Astronomy, University of Leicester, University Road, Leicester, LE1 7RH, UK}


\author[0000-0001-7416-7522]{D.~R. Anderson}
\affiliation{Department of Physics, University of Warwick, Gibbet Hill Road, Coventry CV4 7AL, UK}

\author[0000-0003-0684-7803]{M.~R.Burleigh}
\affiliation{School of Physics and Astronomy, University of Leicester, University Road, Leicester, LE1 7RH, UK}

\author[0000-0001-6472-9122]{L. Raynard}
\affiliation{School of Physics and Astronomy, University of Leicester, University Road, Leicester, LE1 7RH, UK}

\author[0000-0001-6604-5533]{R.~G. West}
\affiliation{Department of Physics, University of Warwick, Gibbet Hill Road, Coventry CV4 7AL, UK}



\begin{abstract}
 \noindent Several studies have documented periodic and quasi-periodic signals from the time series of dMe flare stars and other stellar sources. Such periodic signals, observed within quiescent phases (i.e., devoid of larger-scale microflare or flare activity), range in period from $1-1000$ seconds and hence have been tentatively linked to ubiquitous $p$-mode oscillations generated in the convective layers of the star. As such, most interpretations for the observed periodicities have been framed in terms of magneto-hydrodynamic wave behavior. However, we propose that a series of continuous nanoflares, based upon a power-law distribution, can provide a similar periodic signal in the associated time series. Adapting previous statistical analyses of solar nanoflare signals, we find the first statistical evidence for stellar nanoflare signals embedded within the noise envelope of M-type stellar lightcurves. Employing data collected by the Next Generation Transit Survey (NGTS), we find evidence for stellar nanoflare activity demonstrating a flaring power-law index of $3.25 \pm  0.20 $, alongside a decay timescale of $200 \pm  100$~s. We also find that synthetic time series, consistent with the observations of dMe flare star lightcurves, are capable of producing quasi-periodic signals in the same frequency range as $p$-mode signals, despite being purely comprised of impulsive signatures. Phenomena traditionally considered a consequence of wave behaviour may be described by a number of high frequency but discrete nanoflare energy events. This new physical interpretation presents a novel diagnostic capability, by linking observed periodic signals to given nanoflare model conditions. 
\end{abstract}



\setcounter{table}{0}
\section{Introduction} 
\label{sec:intro}
Magnetic reconnection is a process occurring throughout the outer solar atmosphere, often visible in the form of solar flares. Energies associated with flares express a wide range of magnitudes and frequencies; from very large, but infrequent X-class flares \citep[with X-ray flux exceeding $10^{-4}$~W/m$^2$ at the Earth, or $\sim10^{31}$~ergs per event;][]{Maehara:2015}, down to micro- and nano-flares, each with energies on the order of $10^{-6}$ and $10^{-9}$, respectively, of a typical X-class flare, but with occurrence rates that are orders of magnitude more frequent than the large-scale events. Stellar flares with energies similar to, and exceeding those of our own Sun have also been observed in many observations of stellar sources \citep[e.g.,][]{Lacy:1976, Maehara:2012, Shibayama:2013, Jackman:2018}, predominantly occurring in stars with convective atmospheres, which is required to generate the magnetic fields responsible for reconnection to take place \citep{Pederson:2016}.

The relationship between flare energy and the frequency of occurrence is commonly described by a power-law \citep{Aschwanden:2000}, \edit1{which applies at both low and high flare energies \citep[]{Aschwanden:2019}}. The power-law exponent governs the frequency, $dN/dE$, of flaring events with an associated energy, $E$, through the relationship,  
\begin{equation}
\label{eqn:powerlaw}
    \frac{dN}{dE} ~\sim~ E^{-\alpha} \ ,
\end{equation}
where $\alpha$ represents the power-law index. Low energy solar and stellar flares have long been a topic of wide interest. \edit1{The power-law relation dictates that low-energy flares will be many many times more frequent than larger events.} \citet{Parker:1988} proposed the power-law index is an indicator of the role of magnetic reconnection in maintaining the multi-million degree solar corona, with $\alpha > 2$ allowing low energy (but highly frequent) nano-flares to supply sufficient thermal energy to the outer solar atmosphere to maintain its elevated temperatures. 

Low energy stellar flares have been investigated by a number of authors \citep[e.g.,][to name but a few]{Hudson:1991, Robinson:1995, Robinson:1999, Kashyap:2002, Gudel:2003, gudel:2004, Welsh:2006, Reale:2016}. Much like their solar counterparts, 
there has been no clear consensus on the flaring rates of small-scale stellar flares, with the proposed power-law indices in the aforementioned studies spanning the range $1.5 \le \alpha \le 2.7$.

A review by \cite{gudel:2004} suggested that power-law indices with $\alpha >2$ may be present in M-dwarfs. \citet{Butler:1986} reported the presence of small-scale microflares in dMe flare stars observations that had previously been considered quiescent. Other authors \citep[e.g.,][]{brasseur:2019} have investigated near-ultraviolet (NUV) flare events, with power-laws of $\alpha = 1.72 \pm 0.05$ uncovered. The authors concluded that NUV flare mechanics are governed by the same physical processes as captured in solar events. 
Optical microflare signatures on M-dwarfs have also exhibited short time-scale variability as discussed by \citet{Schmitt:2016}, finding flare rise timescales on the order of seconds, with flare signatures of comparable brightness to its quiescent B-band luminosity. These studies highlight the growing interest in small-scale flare events, and demonstrate the synergy between stellar and solar observational and modeling efforts.

However, there is a gap in the current literature, with few studies investigating the role of nanoflares on other stellar sources. \citet{Falla:1999} examined the production of nanoflare energies in the X-ray emission of RS~CVn systems. The authors concluded that while nanoflares may be produced in these stars, current observational limits would prohibit the direct detection of nanoflare events in the X-ray band. True to the predictions of \citet{Falla:1999}, currently the lowest energy stellar flares that have been directly observed are on the order of $10^{28}$~ergs \citep{Gudel:2002, Benz:2010}, which are orders of magnitude above the traditional range of individual nanoflare energies. It is generally predicted that the flare occurrence rate will be higher on magnetically active stars, such as dMe flare stars \citep{Walkowicz:2011}. As such, nanoflares may be even more frequent on these stellar sources when compared to the Sun, thus producing power-law indices substantially larger than estimates for the solar case. 

Direct observation of solar nanoflares has also remained a challenging endeavor, with their signals lying below the noise floor of current generation instrumentation. As a result, researchers have had to turn their attention to other approaches, such as spectroscopic techniques to compare the scaling between kinetic temperatures and emission measures of
coronal plasma \citep[e.g.,][]{Klimchuk:2001, Sarkar:2008, Sarkar:2009, Bradshaw:2012}, comparisons drawn between EUV and X-ray emission \citep[e.g.,][]{Sakamoto:2008, Vekstein:2009}, or the examination of the time delays between different temperature-sensitive EUV imaging channels \citep[e.g.,][]{Viall:2011,Viall:2012,Viall:2013,Viall:2015, Viall:2016, Viall:2017}. In addition, \citet{Terzo:2011} and \citet{Jess:2014} employed statistical techniques to provide evidence of solar nanoflares. These statistical approaches are further developed in the recent work by \citet{jess:2019}, who infer the presence of nanoflares in a seemingly quiescent solar dataset by comparing intensity fluctuations extracted from high time resolution imaging with those from Monte Carlo synthetic lightcurves designed to replicate the presence of small-scale nanoflare events. \citet{jess:2019} further suggested that similar nanoflare statistical techniques could also be directly applied to high time resolution observations of stellar sources, i.e., to modernize the work of \cite{Audard:1999} and \cite{Kashyap:2002} through the comparisons of intensity fluctuations with nanoflare-specific simulations. 
 
On the contrary to the flare frequencies predicted by the $dN/dE$ power-law relationship, several studies have documented evidence for `periodic' brightness variability through the examination of stellar intensity fluctuations, with periods ranging between $1-1000$~s , \citep{Andrews:1989,Rodriguez:2016,McLaughlin:2018}. These periodic brightenings are of uncertain origin, but are believed to be linked to ubiquitous $p$-mode oscillations or other magnetohydrodynamic (MHD) wave behavior \citep[e.g.,][]{Aschwanden:1999b, Nakariakov:2005, nakariakov:2010, McLaughlin:2018} generated in the convective layers of stars. The link to $p$-mode oscillations is due to a comparable period range ($1-1000$~s), in addition to them being observed during periods of quiescence (i.e., no associated macroscopic flaring signatures).

In a number of publications, \citet{Andrews:1989, Andrews:1990a, Andrews:1990b} examined dMe flare stars across a range of conditions, from immediately after large-scale flare events, to during relatively long periods of quiescence, and found that the dMe flare stars exhibited small periodic brightenings, on a scale of seconds to minutes. The author interpreted these periodic signals as a likely consequence of MHD wave behavior, as the periodic signals were observed during times of quiescence, with no impulsive activity witnessed in the time series. A follow up study by \citet{Andrews:1993} investigated whether flaring events can re-produce signals with $1-1000$~s periodicities, and suggested that while individual small-scale flares may contribute to such signatures, they were unable to provide sufficient evidence to directly link flaring events to the periodic signals.
 
However, flare-related variability giving rise to periodic phenomena has been documented across a range of solar observing sequences. \citet{McLaughlin:2018} discuss self-oscillatory flaring \citep[perhaps due to magnetic dripping, as discussed by][]{nakariakov:2010}, which can produce a periodic signal, despite non-periodic driving. Additionally, \Citet{Arzner:2004} discussed flare clustering, and the relationship between the mean flaring interval and expected count rates. This led \citet{jess:2019} to speculate that small-scale flaring may have a quasi-periodic nature, due in part to the power-law governing its occurrence rates. With this in mind, the superposition of hundreds or thousands of (quasi-) periodic nanoflare signatures each second may give rise to a periodic brightness signal, without any `flare-like' impulsive signatures seen in the corresponding stellar lightcurve. By combining the statistical parameterization techniques developed for solar nanoflare detection with a novel Fourier spectral analysis,  here we investigate stellar nanoflare signals and their potential role in periodic brightenings found in stellar lightcurves.

\section{Observations With NGTS }
The impulsive rise and subsequent decay phase for solar nanoflares are on the order of tens to hundreds of seconds \citep{jess:2019}. Stellar flare decay rates on UV Ceti-type stars are around one order of magnitude shorter than for the Sun, leading to even faster signal evolution on the order of tens of seconds \citep{Gershberg:1975}. As a result, high-frequency resolution and a short temporal cadence are required to fully capture these dynamic signals. \citet{Jackman:2018, Jackman:2019a, Jackman:2019b, Jackman:2019c, Jackman:2020} employed the high cadence of the Next Generation Transit Survey \citep[NGTS;][]{Wheatley:2017} to apply techniques developed for solar flare analysis to stellar flare oscillations, inspiring our use of the NGTS to extend statistical solar nanoflare techniques to stellar lightcurves. The NGTS is a ground-based array of 12 telescopes that scan the sky in the optical domain searching for transiting exoplanet signals, but has also become a platform for stellar flare analyses. The NGTS has  cadence of $\approx$12~s, providing a Nyquist frequency of $\approx$41.6~mHz, with the observations spanning up to hundreds of thousands of frames for a single star. 

When searching for signatures of nanoflare activity we extracted lightcurves for M-type stars. \edit1{ M-dwarf flares have a higher contrast due to their lower quiescent background flux than is typically seen on G and K stars  \citep{Gunther:2020}. This increased contrast is essential to capture nanoflare signals below the noise floor. Additionally, M-star flares have a strong contribution in white-light  \citep{Walkowicz:2011}, ideal when utilizing datasets from optical surveys, i.e., the NGTS. These benefits outweigh an increased photometric noise level, which is itself minimized by leveraging the large number statistics of statistical nanoflare analysis. Finally, as these are flare-active stars , flare occurrence rates will be higher than in `solar-like' stars. This means M-dwarf stars are likely to provide the best conditions for the manifestation of detectable nanoflare signals. }  Specifically, the stars NGTS J030047.1-113651, NGTS J030415.6-103712, and NGTS J031800.1-212036 were chosen as each of these had more than $10^5$ datapoints available for study, hence maximizing the available number statistics for our analyses.

As our scientific analyses revolves around flare-active M-type stars, it was deemed important to also examine non-flare active stars, which can act as a control test to ensure our data analysis techniques are not incorrectly mistaking residual systematic signals as evidence for stellar nanoflares. Since A-type stars are absent of a convective zone, their resulting lack of flare-like behavior provides an ideal set of complementary data products. \edit1{Some recent studies \citep{Balona:2012, Fossati:2018,Balona:2020} do suggest A-stars are capable of flaring, but this has also been disputed \citep{Pederson:2016}. If the observed signals are indeed A-star flares, then only extremely energetic flares have been observed; \citet{Balona:2020} discuss A-star flares with energies in the range $10^{35} - 10^{36}$~ergs, 10 orders of magnitude above traditional nanoflare activity. If only highly energetic events can rise above the high background luminosity on A-stars, this would explain the rarity of A-star flare observations. As such, low-energy nanoflaring would be entirely lost within the lightcurves of these stars due to the minimal contrast invoked, meaning A-type stars would appear quiescent at small-scale flare energies in the NGTS datasets, regardless of their true flaring behavior. This meant A-type stars cannot exhibit a signal consistent with nanoflares.} As such, we examined the A-type stars NGTS J025840.5-120246, NGTS J030958.4-103419, and NGTS J030129.4-110318. 
\edit1{It is important to note that A-type stars have a very different spectral energy distribution to M-dwarf stars, so are not a conventional choice for relative photometric comparison. To ensure robust null testing, we also examined low-activity K stars which have a more comparable SED to M-type stars \citep[i.e., choosing similar spectral types as is standard for photometric comparison, e.g., ][]{Amado:2000}. The low-activity K-type stars were chosen over low-activity M-dwarf stars due to their higher luminosity, leading to decreased low-energy flare contrast when compared to the M-types. While the low-activity K-type stars could theoretically have some weak nanoflaring signature present, it would be minimized compared to the M-types, so this still serves as a valid null test. We used the K2V type stars  NGTS J030000.7-105633 , NGTS J030848.9-112217 , and NGTS J030538.9-114145. These K-stars were low-activity and had no macroscopic flare events in their observed timeseries. } All of the A-type stars , \edit1{K-type stars} , and two of the three M-type stars were obtained from the same observational field (NG0304-1115) and camera (809), hence ensuring consistency across the processed A-, \edit1{K-,} and M-type data sequences. NGTS J031800.1-212036 was from a different field  (NG0313-2230), but had noise statistics, magnitude, and stellar parameters consistent with the other M stars used in the present study.

The magnitudes of the stars employed were comparable (see Table~{\ref{tab:magnitude}}). This was important to ensure the noise statistics were consistent across the stars. The majority of the stars were around mag~13. At this magnitude, the dominant noise source is photon noise \citep[see Figures 3 \& 14 of][]{Wheatley:2017}, with scintillation noise only becoming dominant at the highest frequencies in the data, which are beyond the typical $p$-mode periodicities we are investigating \citep{Osborn:2015}. The A star NGTS J030129.4-110318 was the brightest, with an NGTS magnitude of 11.69. At this magnitude, scintillation became a dominant source of noise. This allowed us to investigate the effect of increased scintillation noise on our analysis techniques. We utilized the stellar parameters from the TESS Input Catalog Version~8 (TIC~V8) \citep{Stassun:2018}, \edit1{along with the initial spectral classification provided via Spectral Energy Distribution (SED) fitting performed by the NGTS pipeline \citep[see section 5.1.1 in][]{Wheatley:2017} to assign the spectral types.} See Table~{\ref{tab:StellarParam}} in the Appendix, for this and other observational parameters \edit1{(i.e., GAIA Source ID, RA, Dec, mass, radius, luminosity, distance, approximate macroscopic flare rates, and the $\log\left(\frac{L_x}{L_{Bol}}\right)$ ratio, where $L_x$ and $L_{Bol}$ are the x-ray and bolometric luminosities, respectively)}.

\edit1{The only M star with an x-ray luminosity measurement was NGTS J030047.1-113651 which had a x-ray flux measurement available from the 4XMM XMM-Newton Serendipitous Source Catalog \citep{Webb:2020}. This corresponded to an x-ray luminosity of $6.47 \times 10^{28}$~ergs{\,}s$^{-1}$. The ratio of x-ray luminosity to the bolometric luminosity is an indication of the activity rate of the star. We find $\log\left(\frac{L_x}{L_{Bol}}\right) = -3.09 \pm 0.21$, which compares to the literature values for a young and active M-type star, with saturated x-ray emission of $\log\left(\frac{L_x}{L_{Bol}}\right) \sim -3 $  \citep{Kastner:2003, Lopez-Santiago:2010}. }

\begin{deluxetable}{ccc}[]
\label{tab:magnitude}
\tablecaption{NGTS magnitudes of the stars used in the study.}
\tablewidth{0pt}
\tablehead{
\colhead{NGTS Identifier} & \colhead{Spectral Type}  & \colhead{NGTS Magnitude}
}
\startdata
NGTS J030047.1-113651 &  M2.5V &  $ 13.23 $ \\
NGTS J030415.6-103712 & M3V  & $ 13.85 $ \\
NGTS J031800.1-212036 &  M2.5V & $ 13.03 $ \\
NGTS J025840.5-120246 & A5V  & $ 13.22 $ \\
NGTS J030958.4-103419 & A5V  & $ 12.55 $ \\
NGTS J030129.4-110318 &  A7V & $ 11.69 $ \\
NGTS J030000.7-105633 & K2V & $13.58 $ \\
NGTS J030848.9-112217 & K2V & $ 13.59 $ \\
NGTS J030538.9-114145 & K2V & $13.57$ \\
\enddata
\end{deluxetable}

\begin{figure*}[!ht]
  \centering
  \includegraphics[trim = 0cm 0cm 0cm 0cm, clip=true, width=\textwidth]{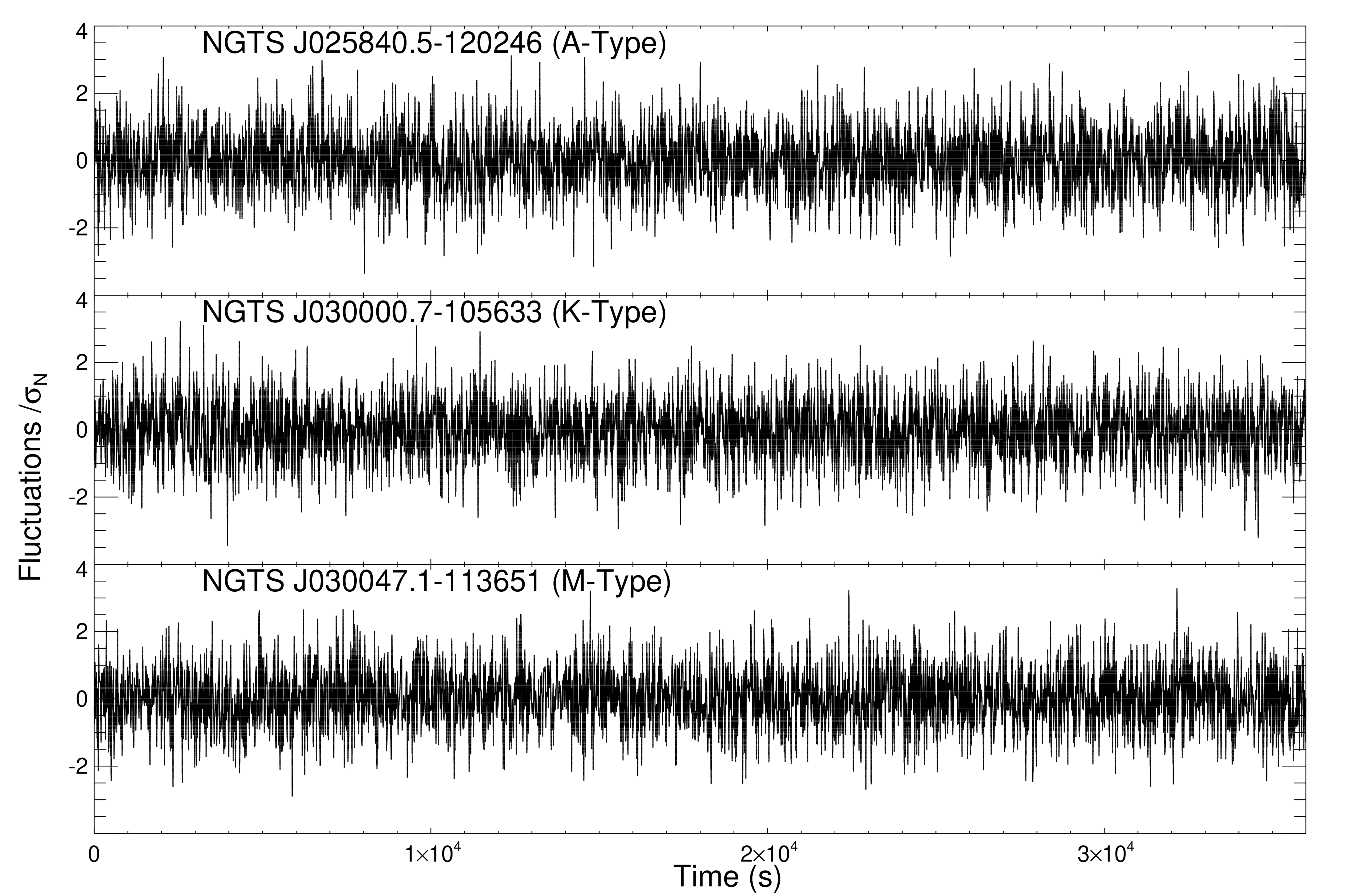}
      \caption{NGTS J025840.5-120246 (A-type; above), \edit1{NGTS J030000.7-105633 (K-type; middle)} and NGTS J030047.1-113651 (M-type; below) lightcurves. These sample lightcurves have been cropped to a 36{\,}000~s interval for clarity, but our analyses utilized the entire time series.  The time interval between successive datapoints is $\sim$ 12~s and the amplitudes have been mean-subtracted and normalized by their respective standard deviations.}
\label{Timeseries}
\end{figure*}
  
The lightcurves were corrected for background and flat-fielded according to the NGTS data reduction pipeline described in \cite{Wheatley:2017}. This pipeline provides a relative error in the flux at each point in the time series. These error bars are affected by cloudy weather and high airmass. Any fluctuations in this error exceeding 1$\sigma$ above the mean value were removed, resulting in $\sim$10\% of each time series being omitted. This removed any data that had statistically significant increases in its associated flux uncertainties, therefore preventing any large flux errors (largely due to  poor seeing conditions) from contaminating the final time series. Next, the lightcurves extracted for each observing sequence were examined for the presence of macroscopic flare signatures, something which occurred in $\sim$0.2\% of the remaining M-type time series (i.e., following the removal of datapoints exceeding 1$\sigma$ in their relative flux errors). To isolate the macroscopic brightenings, each lightcurve was searched for emission signatures exceeding 3$\sigma$ above the mean value, which lasted continually for a minimum of 1~minute (5~datapoints). Based on a normal distribution, the probability of this occurring by chance is {$\lesssim2\times10^{-13}$}, and hence allowed for the robust detection of intensity fluctuations resulting from macroscopic flaring activity. Once the larger scale flare signatures had been identified, they were cut from the time series using an interval of $\pm$5 minutes (25~datapoints) from the first and last detection above the 3$\sigma$ threshold. For consistency, the same processing steps were applied to the A-type \edit1{ and K-type} stellar lightcurves, but no macroscopic brightenings were found for these sources. \edit1{The number of macroscopic flares removed were used to calculate approximate flare rates for the M stars. These  are listed in Table~{\ref{tab:StellarParam}} in the Appendix. The flare rates were of a comparable magnitude for the three M stars, with rates of 0.012, 0.027 and 0.003 flares removed per hour for NGTS J030047.1-113651, NGTS J030415.6-103712, and NGTS J031800.1-212036 respectively. Combining this with the x-ray luminosity of NGTS J030047.1-113651 being that expected for a young and active M-star, we extrapolate that all three M stars are macro-flare active, with roughly comparable activity levels. }

Upon completion of the lightcurve filtering, the lowest number of datapoints remaining was 97{\,}060. To ensure consistency across all subsequent analyses, each of the other \edit1{eight M- K- and A-type} time series were cropped to the same 97{\,}060 datapoints.

\begin{figure*}[!t]
\centering
\includegraphics[ clip=true, angle=0, width=\textwidth]{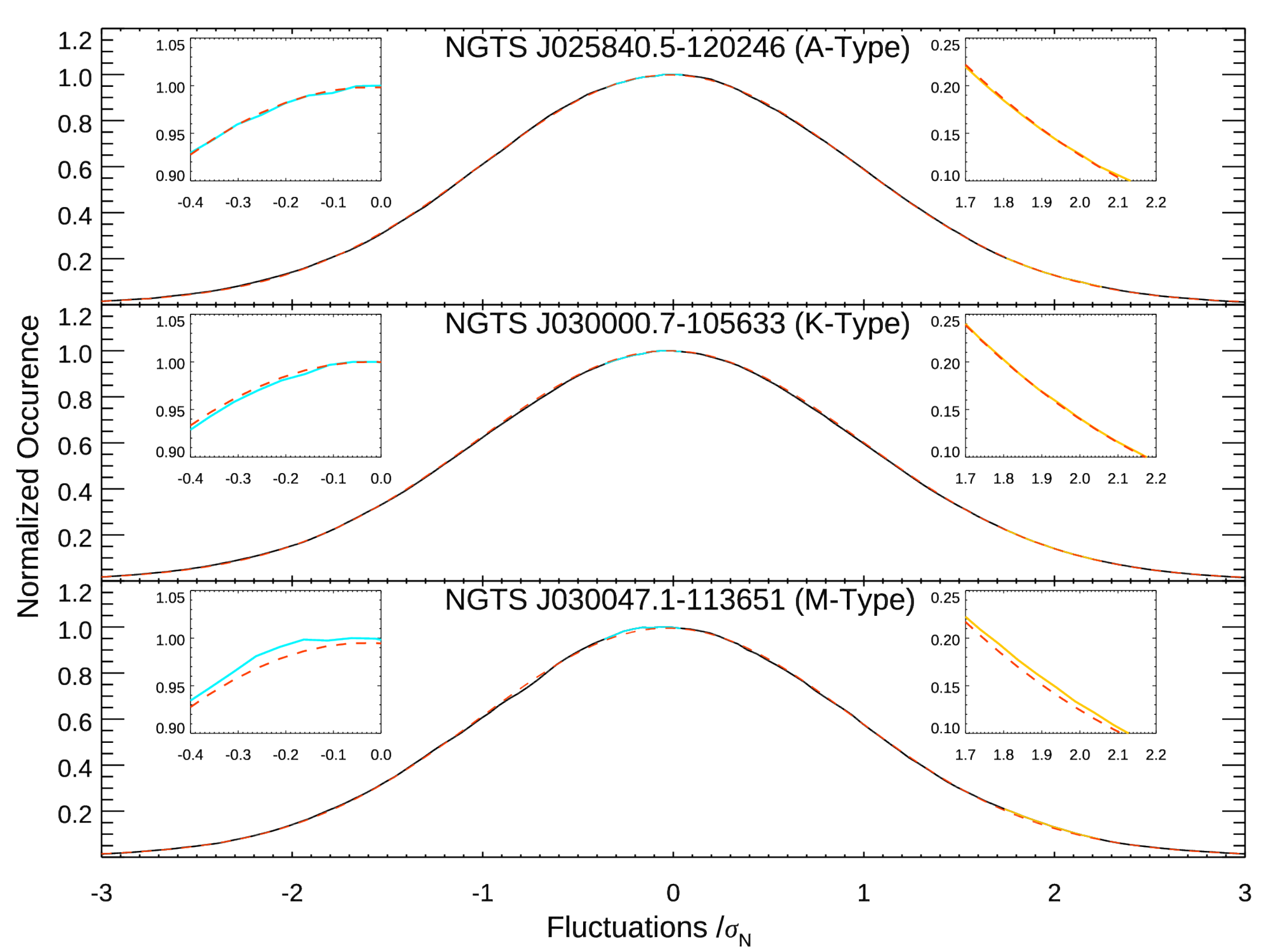}
 \caption{Histograms of intensity fluctuations, each normalized by their respective standard deviations, $\sigma_{N}$, for the NGTS J025840.5-120246 (A-type; above) , \edit1{NGTS J030000.7-105633 (K-type; middle) }, and NGTS J030047.1-113651 (M-type; below) lightcurves. A standardized Gaussian profile is overplotted in each panel using a red dashed line for reference. The M-type distribution has a negative median offset with respect to the Gaussian, in addition to elevated occurrences at $\sim 2~\sigma_{N}$, which is consistent with the statistical signatures of nanoflare activity. On the other hand, the A-type \edit1{ and K-type} intensity fluctuations provide no signatures of flare activity, with the resulting distribution remaining consistent with the presence of photon-based shot noise. Zoomed insets highlight the ranges spanning $-0.4 \le \sigma_{N} \le 0.0$ and $1.7 \le \sigma_{N} \le 2.2$, where M-type negative median offsets and occurrence excesses, respectively, are found. The blue and gold lines display the derived distributions,  \edit1{The M-type exhibited a small dip below the idealized Gaussian at around $-0.90~\sigma_{N}$, which \edit1{are not seen in the A- and K-star}. We believe this is connected to the negative median offset signal, which is causing a consequential dip elsewhere in the statistical distribution, but the exact nature of the signal is unknown.}}
\label{Histogram}
\end{figure*}

Once the macroscopic flare signatures had been extracted from the time series, each of the remaining lightcurves were normalized (night by night) by subtracting a linear line of best fit that was derived from the corresponding time series. Next, the lightcurves were divided by their respective standard deviations, $\sigma_{N}$, providing time series of fluctuations around a common mean that can be readily cross-compared with other star types and data products. This statistical treatment resembles common $Z$-score testing, which is a statistical technique regularly employed in physical and social sciences \citep{Sprinthall:2012}. To ensure that the output data products did not contain any long-term and/or instrumental trends that are not accounted for using the initial preparatory routines, we subsequently detrended these data products using low-order polynomial fits.

\section{Analysis and Discussion}
As documented by \citet{Terzo:2011} and \citet{Jess:2014, jess:2019}, time series commonly referred to as `quiescent’ may in fact contain a wealth of small-scale nanoflare signatures that are embedded within the inherent noise of the photometric signals. It is possible to uncover these signatures through statistical analyses of the intensity fluctuations. We employ the same techniques described by \citet{jess:2019} to attempt to recover nanoflare signatures in our M-dwarf lightcurves.

The dominant source of noise in seemingly `quiescent’ NGTS lightcurves will be shot noise, which follows a Poisson distribution \citep{Wheatley:2017}. The fluctuations will be random, and in the limit of large number statistics, will demonstrate equal numbers of positive and negative fluctuations about the time series mean \citep{Frank:2009}. Therefore, plotting a histogram of the inherent shot noise fluctuations for a truly quiescent time series would produce a symmetric distribution, with the mean and median centered at zero. Any subtle offsets and/or asymmetries to this idealized case may be interpreted as signatures of impulsive events, with subsequent exponential decays, embedded within the noise floor of the lightcurve \citep{Terzo:2011}.

\begin{deluxetable*}{ccccccll}[!t]
\label{tab:stats}
\tablecaption{Characteristics of the intensity fluctuation histograms associated with the A- , K-, and M-type NGTS sources. Note that a standard Gaussian distribution will demonstrate $\zeta=1.73$, hence deviations from this provide an indication of the intensity fluctuation occurrences taking place close to, and far away from the time series mean.} 
\tablewidth{0pt}
\tablehead{
\colhead{NGTS Identifier} & \colhead{GAIA Source ID}  & \colhead{Spectral Type} & \colhead{Datapoints} & \colhead{Median Offset ($\sigma_N$)}  & \colhead{Fisher Skewness} & \colhead{$\zeta$ Ratio} & \colhead{Kurtosis}
}
\startdata
NGTS J030047.1-113651 & 5160579407177989760 & M2.5V  &  $97{\,060}$ & $-0.050 \pm 0.004$ & $0.031 \pm 0.008$ & $1.745 \pm 0.015$ & $0.102 \pm 0.016$  \\
NGTS J030415.6-103712 & 5160771340676667776 & M3V  &  $97{\,060}$ & $-0.050 \pm 0.004$ & $0.009 \pm 0.008$ & $1.783 \pm 0.015$ & $0.130 \pm 0.016$  \\
NGTS J031800.1-212036 & 5099679725858611840 & M2.5V  &  $97{\,060}$ & $-0.049 \pm 0.004$ & $0.041 \pm 0.008$ & $1.761 \pm 0.015$ & $0.169 \pm 0.016$  \\
NGTS J025840.5-120246 & 5160183681775577472 & A5V &  $97{\,060}$ & $0.000 \pm 0.004$ & $-0.032 \pm 0.008$ & $1.766 \pm 0.015$ & $0.133 \pm 0.016$   \\
NGTS J030958.4-103419 & 5165979280580778624 & A5V  &  $97{\,060}$ & $0.000 \pm 0.004$ & $-0.004 \pm 0.008$ & $1.761 \pm 0.015$ & $0.183 \pm 0.016$  \\
NGTS J030129.4-110318 & 5160773569763964416 & A7V  &  $97{\,060}$ & $0.000 \pm 0.004$ & $0.003 \pm 0.008$ & $1.814 \pm 0.015$ & $0.688 \pm 0.016$  \\
NGTS J030000.7-105633 & 5160700765773865600 & K2V  &  $97{\,060}$ & $0.000 \pm 0.004$ & $-0.010 \pm 0.008$ & $1.745 \pm 0.015$ & $0.144 \pm 0.016$  \\
NGTS J030848.9-112217 & 5165722991292368384 & K2V  &  $97{\,060}$ & $0.000 \pm 0.004$ & $-0.008 \pm 0.008$ & $1.723 \pm 0.015$ & $0.119 \pm 0.016$  \\
NGTS J030538.9-114145 & 5159884962505997184 & K2V  &  $97{\,060}$ & $0.000 \pm 0.004$ & $-0.013 \pm 0.008$ & $1.723 \pm 0.015$ & $0.118 \pm 0.016$  \\
\enddata
\end{deluxetable*}

As discussed by \citet{jess:2019}, nanoflares give rise to two distinct signals in the resulting intensity fluctuation histograms. The first is a negative median offset, whereby the median value of the histogram is $< 0~\sigma_N$. This is a characteristic signal associated with an exponentially decaying signature, i.e., the decay phase following an impulsive deposition of energy occurs over a longer timescale, hence providing more fluctuations that are beneath the elevated signal mean caused by the impulsive event. The second signature is an excess of fluctuations at $\sim 2~\sigma_N$, which is caused by the impulsive nature of the nanoflare intensity rises, and gives rise to an asymmetric distribution that can be benchmarked using Fisher skewness coefficients. As the evolution of a nanoflare produces an almost discontinuous increase in the lightcurve intensity, a distinct positive peak manifests in the resulting histogram of intensity fluctuations. Therefore, a seemingly quiescent lightcurve exhibiting both of these signals is a strong candidate to contain embedded nanoflare signatures. Additionally, we benchmark the shape and widths of the distributions through calculation of the histogram kurtosis values, in addition to the ratio of its full-width at eighth-maximum to that of its full-width at half-maximum (i.e., FW$\frac{1}{8}$M-to-FWHM ratio), which is defined as `$\zeta$' for simplicity \citep[see][for a more thorough overview of this key statistical parameter]{jess:2019}. Note that a standard Gaussian distribution will have $\zeta=1.73$, hence deviations from this provide an indication of the intensity fluctuation occurrences taking place close to, and far away from the time series mean.

\subsection{NGTS Datasets}
Figure~{\ref{Timeseries}} displays sample lightcurves, cropped to a 36{\,}000~s interval, where the intensity fluctuations are normalized about their respective means and standard deviations. Figure~{\ref{Histogram}} displays the intensity fluctuation histograms for both the example A- \edit1{K-} and M-type stellar sources NGTS J025840.5-120246, \edit1{NGTS J030000.7-105633}, and NGTS J030047.1-113651, respectively. As expected, the non-flare active A-type star \edit1{and low-activity K-star} show little variation from the standardized Gaussian distribution (dashed red lines in Fig.~{\ref{Histogram}}), with median offsets of $0.000\pm0.004~\sigma_{N}$ and no visible excess at ${\sim+2~\sigma_{N}}$. This suggests that the \edit1{A-type and K-type stars have} no embedded nanoflare characteristics, and therefore reiterates \edit1{their} importance as a control test for subsequent M-type star analysis. On the other hand, the M-type star displays both of the characteristic nanoflare signatures, with a negative median offset equal to $-0.050\pm0.004~\sigma_{N}$, and a visible occurrence excess at $\sim +2~\sigma_N$, culminating in an associated positive Fisher skewness value of $0.031\pm0.008$ that is above the expectations of a pure Gaussian distribution. 

The other candidate stars exhibited consistent signals, with the M-type stars showing histogram signatures consistent with nanoflare activity, while the A- \edit1{and K-}type stars showed no indication of impulsive behavior beneath the noise floor. A-type star NGTS J030129.4-110318 exhibited a small positive skew of $0.003 \pm 0.008$, but the associated uncertainty makes this less definitive when compared to the positive skewness values exceeding $0.040$ for some M-type sources. Furthermore, NGTS J030129.4-110318 also demonstrated zero median offset, remaining inconsistent with a distribution comprised of impulsive events followed by gradually decaying tails. This star had a much larger deviation from Gaussian statistics, evidenced by a kurtosis value of $0.688 \pm 0.016$, and $\zeta$ ratio of $1.814 \pm 0.015$.  This deviation from Gaussian statistics is due to the increased brightness of this star (see Table~{\ref{tab:magnitude}}) compared to the other candidates, resulting in scintillation becoming a more significant noise source \citep{Osborn:2015, Wheatley:2017}. It is important to note that while the scintillation noise produces statistics offset from a Gaussian, it is still distinct from the characteristic signatures of nanoflaring. This highlights the robustness of the statistical nanoflare analysis. The characteristics derived for all 9 stellar sources are documented in Table~{\ref{tab:stats}}. 
\edit1{The M-type stars exhibited a small dip below the Gaussian around $-0.90~\sigma_{N}$, which was not seen in the A- \edit1{or K-}stars. We believe this is connected to the negative median offset signal, which is causing a dip elsewhere in the statistical distribution, but the exact nature of the signal is unknown. Future investigation could uncover the source of this dip, and potentially use it as a further diagnostic. }

\begin{figure*}
   \centering
   \includegraphics[ width=\hsize]{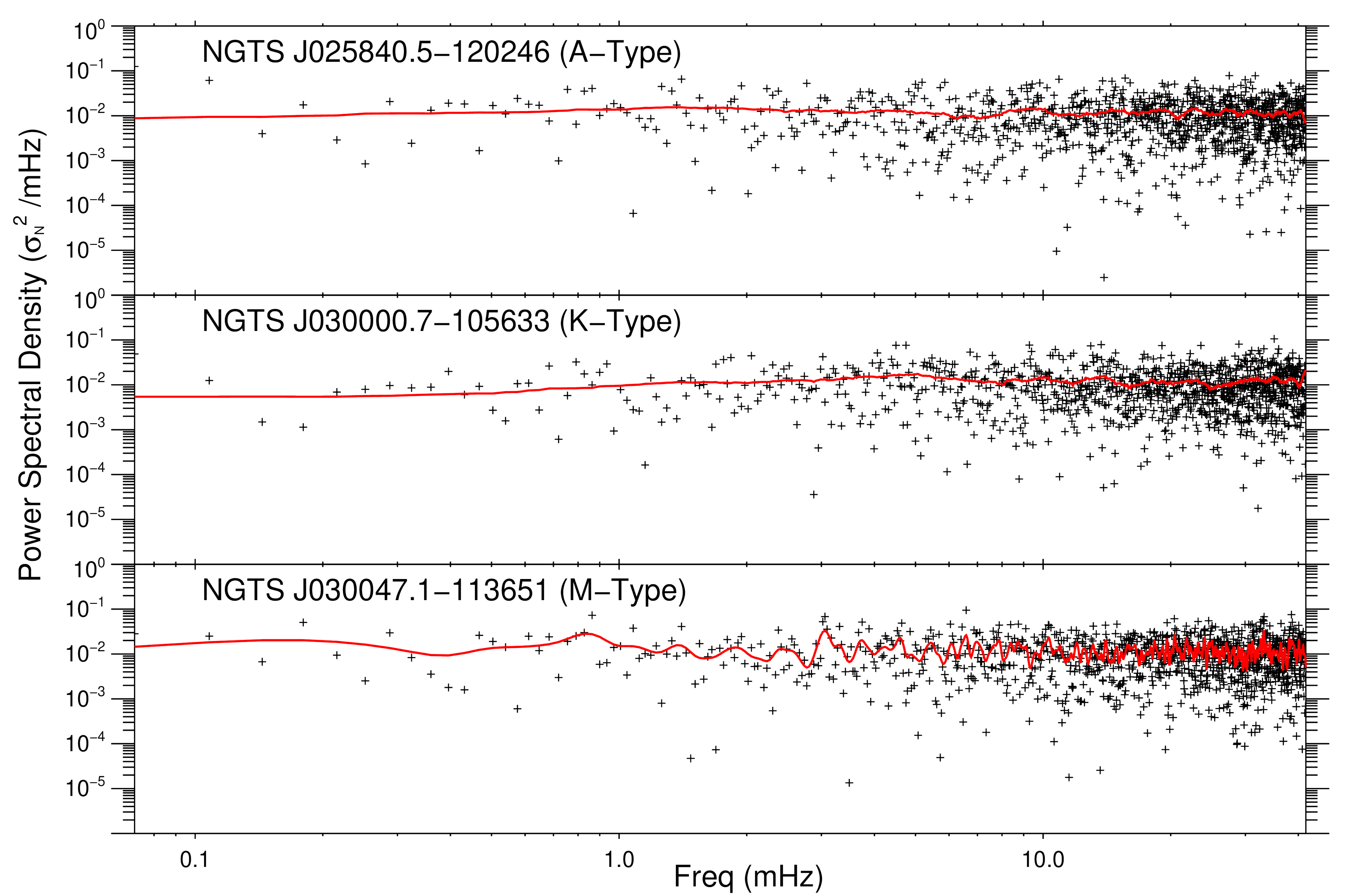}
      \caption{The Fourier power spectral densities (PSDs) for example A- \edit1{K-} and M-type stellar sources NGTS J025840.5-120246 (upper panel)  \edit1{NGTS J030000.7-105633 (middle panel)} and NGTS J030047.1-113651 (lower panel), respectively, displayed in normalized units of $\sigma_{N}^{2}/\text{mHz}$. The crosses in each panel depict the individual power values as a function of frequency, while the solid red line reveals a trendline calculated over $\pm$6 frequency elements ($\pm$0.427~mHz). It can be seen that the\edit1{ A- and K-type  PSD are relatively flat}, with no clear power enhancements,\edit1{ apart from slight enhancement in the K-type star, in the range of $1-10$~mHz, indicative of the expected p-mode oscillations seen in Solar-like stars}. Contrarily, the M-type PSD has a primary power peak at $\approx$0.8~mHz, followed by decreasing spectral power exhibiting a spectral slope of $\beta = -0.30 \pm 0.05$, followed by numerous power peaks in the range of $3-10$~mHz, consistent with previous links to stellar $p$-mode spectra.}
         \label{PSD_NGTS}
   \end{figure*}

Employing the high time resolution and long duration imaging sequences of the NGTS data products has enabled us to provide the first tentative evidence of nanoflares occurring on stellar sources (see, e.g., Fig.~{\ref{Histogram}} and Table~{\ref{tab:stats}}). However, while the statistical signatures derived for the NGTS M-type lightcurves resemble those expected for nanoflare activity, they do not provide any indication of the specific underlying plasma conditions at work.

As previously demonstrated by \citet{Andrews:1989}, \citet{Rodriguez:2016}, and \citet{McLaughlin:2018}, small-scale brightenings -- here hypothesized to be the result of nanoflare activity -- often give rise to periodic signatures in the corresponding lightcurves. This has also been observed in the case of small-scale solar activity \citep{Terzo:2011}. To investigate the manifestation of periodicities in the stellar lightcurves, time series were extracted for each star that contained the maximal number of successive frames, where no breaks resulting from problematic flux calibrations, macroscopic flare events, or day/night cycles were present, i.e., the longest consecutive number of frames consistent across the \edit1{9} stars. The lowest number of viable consecutive frames was 2316 from M-type star NGTS J031800.1-212036. As such, each of the remaining five lightcurves were cropped to an identical 2316~datapoints ($\approx 27{\,800}$~s duration) so that the final A- \edit1{,K- ,} and M-type time series had identical lengths, helping to ensure consistency between both the Nyquist frequency and frequency resolution in the subsequent analyses.

\begin{deluxetable*}{cccccc}[!t]
\label{tab:PSDStats}
\tablecaption{Characteristics of the PSDs associated with the M-type NGTS time series.}
\tablewidth{0pt}
\tablehead{
\colhead{NGTS Identifier} & \colhead{Spectral Type} & \colhead{Number of datapoints} & \colhead{Gradient}   & \colhead{Turning Point (mHz)} & \colhead{Peak Frequency (mHz)}
}
\startdata
NGTS J030047.1-113651 & M2.5V  & $2316$  & $-0.30\pm 0.05$  & $0.81 \pm 0.04$ & $2.90 \pm 0.04 $ \\
NGTS J030415.6-103712 & M3V  & $2316$ & $-0.28 \pm 0.05$ &  $0.90 \pm 0.04$ & $2.58 \pm 0.04$  \\
NGTS J031800.1-212036 & M2.5V  & $2316$ & $-0.26 \pm 0.05$  & $0.64 \pm 0.04$ & $3.09 \pm 0.04$  \\
\enddata
\end{deluxetable*}

Each of the six extracted NGTS lightcurves were passed through a Fast Fourier Transform (FFT) to determine whether power exists at frequencies synonymous with a typical $p$-mode spectrum \citep[often in the range of $1-1000$~s;][]{Kjeldsen1995, Guenther2008, Handler2013, DiMauro2016}. The input data resulted in a Nyquist frequency of $\approx$41.6~mHz being complemented by a frequency resolution, $\Delta{f} = 0.0356$~mHz, in the corresponding FFTs. However, it must be pointed out that a strictly periodic wave signal would not manifest as median offsets and/or asymmetries in the fluctuation histograms documented in Figure~{\ref{Histogram}}, since the evolution of a purely sinusoidal wave signal is symmetric about its given mean. The resulting Fourier power spectra were transformed into power spectral densities (PSDs) following the methods defined by \citet{Welch:1961} and \citet{Vaughan:2013}.

Following the generation of PSDs from the \edit1{nine} NGTS lightcurves, we find that the A- \edit1{, K- ,} and M-type sources exhibit consistent and distinct features in their corresponding PSDs, with examples depicted in Figure~{\ref{PSD_NGTS}}. The upper \edit1{, middle, and} lower panels of Figure~{\ref{PSD_NGTS}} display the PSDs for the A-, \edit1{K-,} and M-type stars NGTS J025840.5-120246, \edit1{NGTS J030000.7-105633 ,} and NGTS J030047.1-113651, respectively. In each panel, the crosses represent the individual frequency-dependent power, while the solid red line depicts a trendline created using a $\pm$6 frequency element ($\pm$0.427~mHz) smoothing. It can be seen from the solid red lines in Figure~{\ref{PSD_NGTS}} that the A-type \edit1{and K-type} spectra are relatively flat across all frequencies with no evidence of distinct peak frequencies. \edit1{The K-type does show some slight power enhancement between $\approx 1-10$~mHz, consistent with stellar $p$-mode oscillations, as have been previously observed in K-type Solar-like stars \citep[e.g.][]{Chaplin:2009} }. The M-type PSD exhibits more pronounced fluctuations across the frequency domain. In the lower panel of Figure~{\ref{PSD_NGTS}}, the solid red line highlights the presence of a primary power peak at $\approx$0.8~mHz, followed by a gradual decline in power as the frequency increases. This reduction in power, as a function of frequency, can be represented by a spectral slope, $\beta$, following the form $f^{\beta}$. In the lower panel of Figure~{\ref{PSD_NGTS}}, the spectral slope is calculated to be $\beta = -0.30 \pm 0.05$. For each M-type star the position of the primary peak, and its associated spectral slope, were calculated. The primary peaks (or `Turning Point') were found in the range of $0.6-0.9$~mHz, with the corresponding spectral slopes calculated to span $-0.30 \leq \beta \leq -0.26$. Once the spectral slopes had been calculated, they were subsequently subtracted from each PSD to better highlight power fluctuations above the background level \citep[similar to the processing undertaken by][]{Krishna2017}. Following the detrending of the PSDs, the frequency demonstrating maximal power above the background was subsequently extracted, and found to reside in the range of $2.58-3.09$~mHz for the M-type stellar sources, which is consistent with previous interpretations related to the presence of $p$-mode oscillations \citep{Andrews:1989, Andrews:1990a, Andrews:1990b}. The specific characteristics derived from the M-type PSDs are displayed in Table~{\ref{tab:PSDStats}}.\edit1{The Figure~{\ref{PSD_ALL}} in Appendix B plots all the PSD trendlines on one plot for clarity.}

With the lightcurve intensity fluctuations statistically benchmarked, and the corresponding power spectra uncovered, we now generate Monte Carlo nanoflare simulations that have been tailored for stellar sources. This will enable direct comparisons to be made between the observed and simulated time series (for both the statistical fluctuations and the power spectra features), which will help quantify the specific plasma parameters at work in each of the stellar sources. \edit1{The modeled time series will be cropped to the same length as the NGTS time series, i.e., 97{\,}060 data points (at a cadence of $ \sim 12 s $) for the statistical analysis, and 2316 data points for the PSDs, to ensure consistency in their number statistics.}

\subsection{Stellar simulations}
We adapt the Monte Carlo simulations described by \citet{jess:2019} to synthesize the intensity time series expected for a broad range of initial plasma conditions. The adaption process first necessitated altering the `area' over which the simulations took place. In the work of \citet{jess:2019}, a two-dimensional image was generated to simulate data acquired by the Atmospheric Imaging Assembly \citep{Lemen2012} onboard the Solar Dynamics Observatory \citep{Pesnell2012}, where the pixels had an area of approximately $10^{15}$~cm$^{2}$. However, considering our one-dimensional lightcurves contain no resolvable spatial information of the stellar sources, we need to increase the modeled area to represent the entire Earth-facing surface area of the star. This meant setting the pixel area to around $10^{21}$~cm$^{2}$, or approximately 10\% of the surface area of the Sun, which corresponds to the surface area of a typical M-dwarf stellar source \citep{Reid:2005}. Note that we use the entire Earth-facing surface area. While larger flares require specific high energy magnetic conditions (e.g., large-scale spots that may only cover a small proportion of the stellar surface and be more aligned with the stellar equator), it is expected that nanoflares can effectively occur anywhere across the stellar atmosphere, requiring only small-scale magnetic activity to trigger them. This area, along with an exposure time (10~s) and final cadence (12~s) matched to the NGTS observations, was used to re-compute the number of flaring events expected \citep[following Equation~{\ref{eqn:powerlaw}} and the work by][]{Aschwanden:2000, Parnell:2000}, for a given power-law index, $\alpha$, and across a specific time interval. \edit1{The quiescent flux of the M-stars was used to generate the underlying Poisson noise in the flare models and the nanoflare energies were then calibrated to this noise level, following the steps taken in \citet{jess:2019} .}
 
The flare energies included in our model spanned $10^{22} - 10^{25}$~ergs, placing them within the energy regime synonymous with solar nanoflares. \edit1{dMe flare stars are not `solar-like'; } arguably, the energy span of M-type stellar nanoflares may be orders of magnitude larger than for the solar case, due to the increased flare energies associated with M-type stars. However, various authors \citep{Falla:1999, Robinson:1999, Gudel:2002} have applied the solar energy span derived by \citet{Aschwanden:2000} directly to stellar investigations, hence we follow the same convention for consistency. The conversion of flare energies to peak detector counts, $DN$, is performed via a direct one-to-one scaling relationship. According to \citet{Yang:2017}, the flare energy is linear with area, which is linear with flux, assuming a constant black-body emission temperature. As flares emit primarily in optical and UV wavelengths \citep{Neidig:1989, Woods:2006, Schmitt:2016}, whitelight observations are likely to capture the resulting nanoflare emission \citep{Kretzschmar2011}, \edit1{particularly for M-dwarf flares which emit strongly in white light \citep{Walkowicz:2011}}, resulting in a $DN \propto E$ relationship. This is similar to the pulse-heating model proposed by \citet{jess:2019}, whereby $DN \propto E^{4/3}$. For the energy range relevant to nanoflares (i.e., spanning only 3 orders-of-magnitude; $10^{22} - 10^{25}$~ergs), the differences between the linear scaling and pulse-heated models is relatively small. However, if accurate modeling and replication of full-scale flaring events (i.e., $10^{22} - 10^{31}$~ergs) is required, then more precise whitelight emission models would need to be developed. \citep{Prochazka2018}.

\begin{figure*}[!t]
  \centering
  \includegraphics[trim = 0cm 0cm 0cm 2.5cm, clip=true, width=\textwidth]{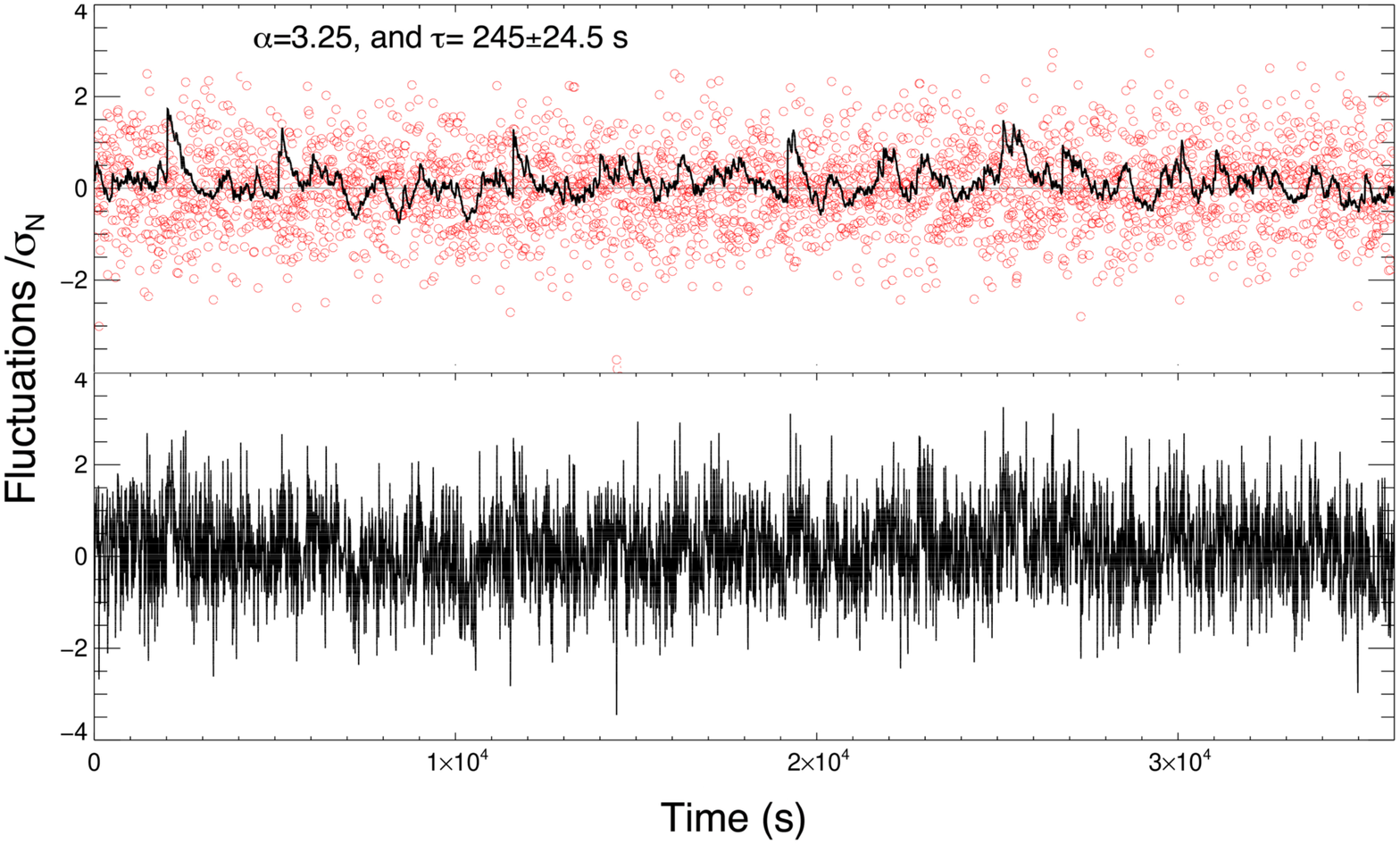}
      \caption{The generation of flare signals according to a power-law relationship, where the power-law exponent is $\alpha=3.25$ and the decay timescale for each event is $245\pm24.5$~s. The superposition of all modeled flare signatures is displayed using the solid black line (upper panel). Red dots represent the shot noise distribution modeled using Poisson statistics. The time interval between successive datapoints is 12~s and the amplitude has been mean-subtracted and normalized by its standard deviation. The lower panel reveals the superposition of the synthetic flaring lightcurve with the Poisson noise model, highlighting the difficulty extracting visual signatures of small-scale flaring events from both synthetic and observational time series. This final time series is comparable to the observed time series, e.g.  Fig.~{\ref{Timeseries}} }
\label{SeqFlares}
\end{figure*}

Due to the large spatial integration ($\approx 10^{21}$~cm$^{2}$), the simulations are more computationally intensive than described by \citet{jess:2019}. As a result of integrating over the entire stellar disk, the generation and superposition of hundreds of thousands of independent nanoflare events becomes a more time consuming endeavor, requiring approximately 300~s on a 2.90~GHz Intel Xeon processor to generate a synthetic NGTS time series incorporating $97{\,060}$ individual frames ($\sim$13.5 continuous days of data at a cadence of 12~s). An example depicting the generation of a synthetic NGTS lightcurve is shown in Figure~{\ref{SeqFlares}}. Here, the lightcurve is cropped to a 36{\,}000~s interval to more clearly reveal its constituent components. The upper panel of Figure~{\ref{SeqFlares}} displays (black line) modeled flaring events  using a power-law index $\alpha = 3.25$ and a decay timescale (i.e., reflecting the $e$-folding time of the flare decays) of $\tau = 245\pm24.5$~s. Note that the decay timescale varies by $\pm$10\% (i.e., $\tau = 245\pm24.5$~s) to allow for subtle variations in the mechanisms responsible for cooling in the immediate aftermath of the flaring events \citep{Antiochos1978}. This nanoflare time series is the superposition of individually generated flare events.
The red dots represent the background shot noise, which follows a Poisson distribution. According to the limits of large number statistics, this Poisson profile will transform into a Gaussian distribution, with $\approx$68.3\%, $\approx$95.5\%, and $\approx$99.7\% of the noise fluctuations contained within the invervals of $\pm 1\sigma_{N}$, $\pm 2\sigma_{N}$, and $\pm 3\sigma_{N}$, respectively. It is visible from the upper panel of Figure~{\ref{SeqFlares}} that even larger flaring events, e.g. occurring at $\sim$200~s and $\sim$500~s, are contained within the noise envelope. Once the shot noise contributions have been added to the synthetic flaring signals, the resulting time series (lower panel of Fig.~{\ref{SeqFlares}}) mimics very closely typical stellar lightcurves (i.e. the NGTS lightcurves in  Fig.~{\ref{Timeseries}}), with the original nanoflare signal now indiscernible from the embedded noise.

Figure~{\ref{SeqFlares}} documents the steps taken to generate a synthetic lightcurve for a specific power-law index ($\alpha = 3.25$) and $e$-folding timescale ($\tau = 245\pm24.5$~s). However, in order to more accurately constrain our observational findings using our synthesized models required us to repeat the processing steps documented in Figure~{\ref{SeqFlares}} using a dense grid of nanoflare input parameters. Specifically, power-law indices spanning $1 \le \alpha \le 4$ (in intervals of $0.05$) and $e$-folding timescales ranging across $5 \le \tau \le 500$~s \citep[in steps of $5$~s, consistent with previous estimations for solar nanoflares;][]{Terzo:2011, Jess:2014} were employed. This produced 6100 final synthetic NGTS lightcurves, each with 97{\,}060 datapoints to remain consistent with the observational NGTS time series, ensuring identical number statistics and allowing direct comparisons to be made between the observations and simulations.

\subsection{Comparing Simulation to Observation}

\begin{figure*}[!t]
\centering
\includegraphics[ clip=true, width=\textwidth]{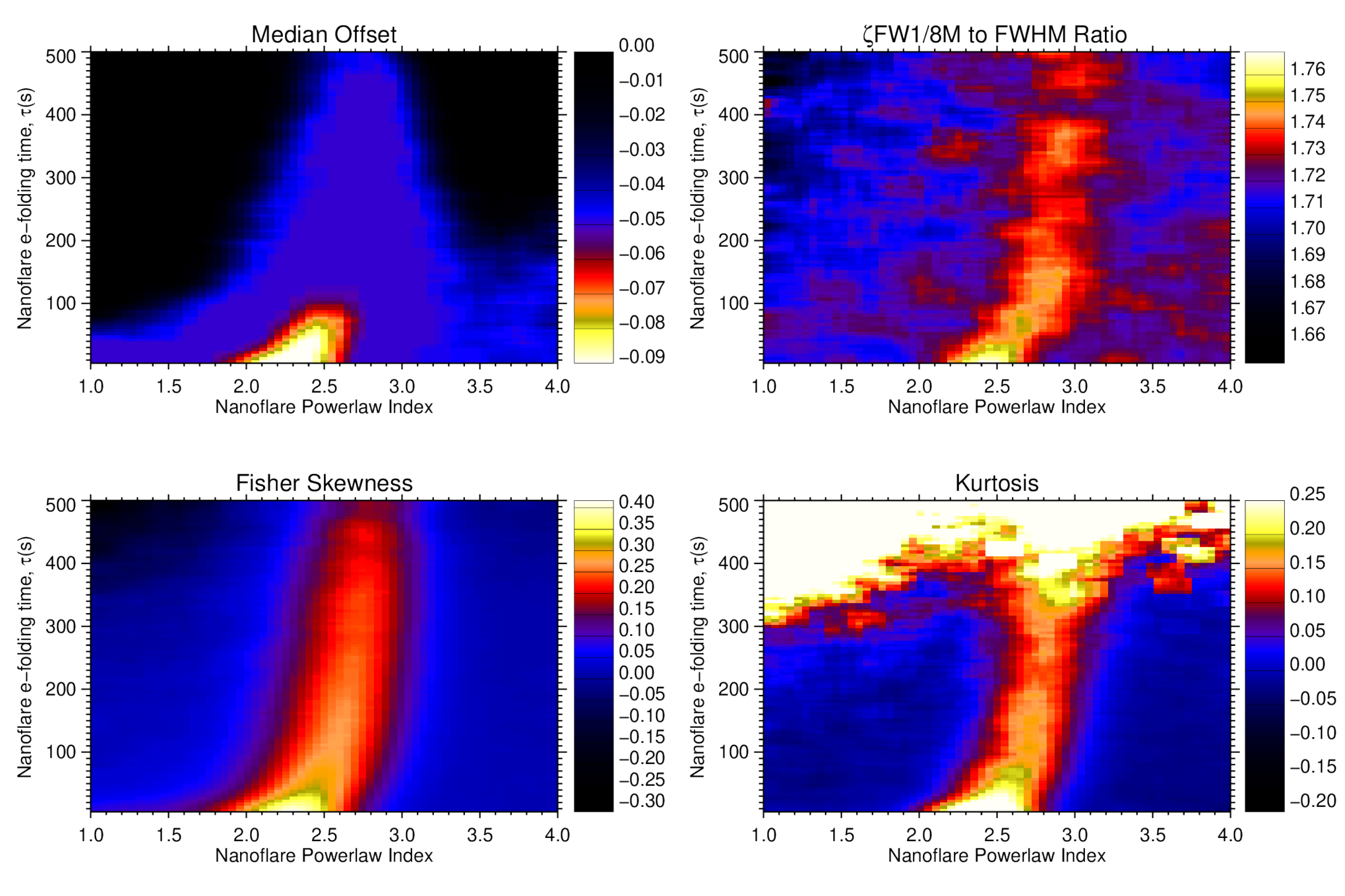}
 \caption{The median offset (upper-left), $\zeta$ (FW$\frac{1}{8}$M-to-FWHM ratio; upper-right), Fisher skewness (lower-left), and kurtosis (lower-right) characteristics extracted from the synthetic intensity fluctuation distributions as a function of the employed power-law index, $\alpha$, and the decay timescale, $\tau$. A negative median offset and positive Fisher skewness values support a wide range of flare conditions. \edit1{The observational statistical characteristics (see Table~{\ref{tab:stats}} and Figure~{\ref{Histogram}}) compare to the modeled statistical distributions with overlapping parameters corresponding to $\alpha=3.25\pm0.15$ and $\tau=200\pm100$~s, in addition to $\alpha=2.00\pm0.15$ and $\tau=200\pm100$~s.} } 
\label{Stat_Map}
\end{figure*}

Each synthetically generated lightcurve was treated in an identical manner to that of the NGTS observations, whereby each of the 6100 simulated time series were detrended and normalized by their respective standard deviations, before generating their intensity fluctuation distributions and subjecting them to FFT analyses. It must be noted that there were no instances in any of the 6100 simulated time series where a sequence of 5 successive time steps exceeded $+3 \sigma_{N}$ above the mean, hence highlighting the consistency between the simulated lightcurves and the final time series extracted from the NGTS observations. 

First, to compare the observational intensity fluctuation distributions depicted in Figure~{\ref{Histogram}} to those extracted from the dense grid of simulation input parameters, we generated a number of statistical maps (Fig.~{\ref{Stat_Map}}) where the parameter values extracted from the intensity fluctuation histograms are displayed as a function of the power-law index, $\alpha$, and the corresponding decay timescale, $\tau$. These statistical benchmarks are the same as those calculated for the NGTS stars in Table~{\ref{tab:stats}}, only now graphically displayed in a two-dimensional format to aid visual clarity.

\begin{figure*}
   \centering
   \includegraphics[ width=\hsize]{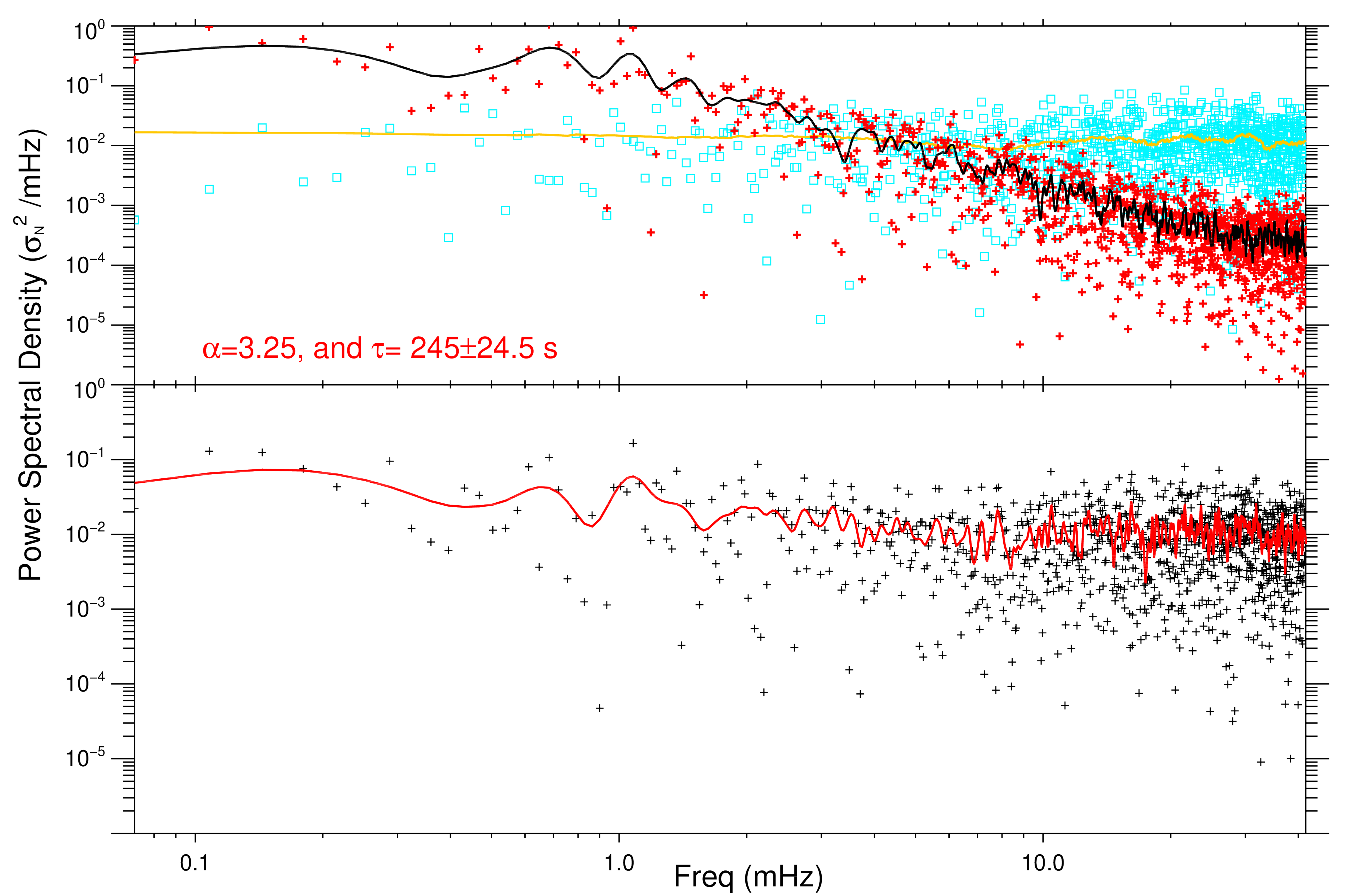}
      \caption{The Fourier power spectral densities, displayed in units of $\sigma_{N}^{2}/\text{mHz}$, corresponding to a power-law index $\alpha=3.25$ and a flare decay timescale $\tau= 245 \pm 24.5$~s. The upper panel depicts the nanoflare and shot noise PSDs as red crosses and blue squares, respectively. The solid black and gold lines represent trendlines for the nanoflare and shot noise profiles, respectively, computed over $\pm$6 frequency elements ($\pm$0.427~mHz). The lower panel displays the PSD of the final synthetic time series, where the nanoflare signal is embedded within the synthetic noise floor. The synthetic PSDs corresponding to nanoflare activity are remarkably similar to those for the NGTS M-type stellar sources shown in the bottom panel of Figure~{\ref{PSD_NGTS}}.}    
\label{PSD_SIM}
\end{figure*}

The measured output parameters depicted in Figure~{\ref{Stat_Map}} allows us to cross-correlate the observational signatures to those synthetically generated via the Monte Carlo modeling work, hence allowing us to estimate the specific plasma conditions (i.e. the $\alpha$ and $\tau$ values) responsible for the observational signatures. Importantly, the synthetic stellar lightcurves are consistent with those expected from solar modeling efforts \citep{Terzo:2011, Jess:2014, jess:2019}, whereby a negative median offset is coupled with an increase in the Fisher skewness value. From Figure~{\ref{Stat_Map}} it can be seen that the majority of nanoflare conditions produce a negative median offset and positive Fisher skewness in the resulting statistical intensity fluctuation distribution, despite the presence of seemingly quiescent lightcurves (see, e.g., the lower panel of Fig.~{\ref{SeqFlares}}). \edit1{Similar dips below the idealized Gaussian at approximately $-0.90 \sigma_{N}$ (as were seen in the M-type stars) were exhibited in the simulations, suggesting these are linked to the embedded nanoflare signals.}

When comparing the intensity fluctuation statistical outputs for the M-type stars to those derived from the Monte Carlo simulations, we found overlap in the median offset, Fisher skewness, kurtosis, and $\zeta$ ratio corresponding to two distinct plasma conditions governed by the flare power-law index, $\alpha$, and the associated decay timescale, $\tau$. The first set of self-similar parameters corresponded to $\alpha=3.25\pm0.15$ and $\tau=200\pm100$~s, while the second set of parameters consisted of $\alpha=2.00\pm0.15$ and $\tau=200\pm100$~s. These values highlight the fact that the observational M-type NGTS lightcurves show remarkable agreement with the statistical signals derived from Monte Carlo synthetic lightcurves consisting of nothing but nanoflare signals embedded in characteristic shot noise. Contrarily, the A-type \edit1{ and K-type} stellar parameters do not map consistently onto the statistical parameters depicted in Figure~{\ref{Stat_Map}}, reiterating our interpretation that the A-type \edit1{ and K-type} sources do not exhibit nanoflare signatures.  

In order to further examine the link between nanoflare activity and periodic variability in the synthetic lightcurves, we generated PSDs for each of the 6100 simulated time series, which could then be compared directly with the PSD features found in the NGTS observations. To remain consistent with the observational PSDs depicted in Figure~{\ref{PSD_NGTS}}, we cropped the synthetic time series to 2316~datapoints to ensure the frequency resolution was maintained at $\Delta{f}=0.0356$~mHz. As the comparison between the observed and modeled intensity fluctuation distributions revealed a self-similar set of statistical parameters corresponding to a power-law index $\alpha=3.25\pm0.15$ and a decay timescale $\tau=200\pm100$~s, we provide example PSDs for $\alpha=3.25$ and $\tau = 245\pm24.5$~s in Figure~{\ref{PSD_SIM}}. Such Fourier analysis offers an additional paramaterization of the nanoflare signal, allowing us to resolve any ambiguities arising through examination of the statistical signatures alone.

\begin{figure*}[!t]
  \centering
  \includegraphics[ clip=true, width=\textwidth, angle=0]{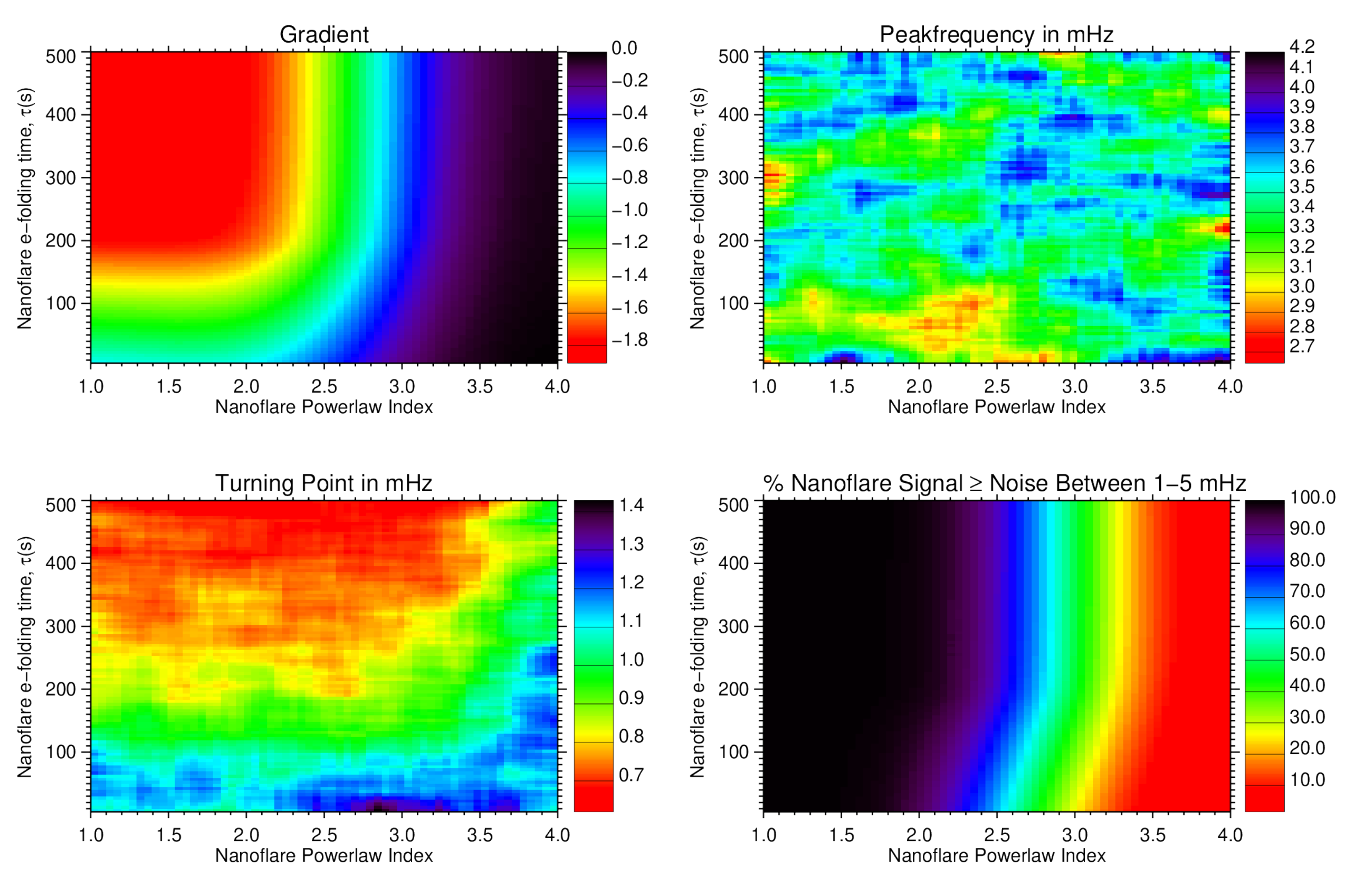}
      \caption{The primary peak frequencies (lower-left), spectral slopes (upper-left), dominant frequencies following detrending (upper-right), and the percentage of nanoflare power above the noise floor in the range of $1-5$~mHz (lower-right), displayed as a function of the power-law index, $\alpha$, and the decay timescale, $\tau$, used to generate the synthetic time series. \edit1{The observational PSD characteristics (see Table~{\ref{tab:PSDStats}} and Figure~{\ref{PSD_NGTS}}) compare to the modeled PSDs in the range of $\alpha  = 3.3 \pm  0.2 $ and $\tau = 200 \pm  100$~s.}}
         \label{Heat Map}
  \end{figure*}
  
The upper panel of Figure~{\ref{PSD_SIM}} shows the corresponding PSDs for both the raw nanoflare (red crosses) and Poisson-based shot noise (blue squares) signals. The solid black and gold lines in the upper panel of Figure~{\ref{PSD_SIM}} depict the trendlines for the nanoflare and shot noise signals, respectively, established over $\pm$6 frequency elements ($\pm$0.427~mHz).  It can be seen that at lower frequencies ($\lesssim5$~mHz) the nanoflare signal dominates over the corresponding noise profile, while at higher frequencies the noise becomes dominant and begins to mask the frequency-dependent signals of nanoflare activity. The lower panel of Figure~{\ref{PSD_SIM}} displays the PSDs extracted from the final simulated lightcurve, where the nanoflare signal has been embedded within the synthetic noise profile. To remain consistent with the lower panel of Figure~{\ref{PSD_NGTS}}, the black crosses represent the individual frequency-dependent power measurements, while the solid red line depicts a trendline established over $\pm$6 frequency elements ($\pm$0.427~mHz). The similarities between the lower panels of Figures~{\ref{PSD_NGTS}} \& {\ref{PSD_SIM}} are remarkable, exhibiting similar primary power peaks at $\sim$1~mHz, followed by a decrease in spectral power with increasing frequency, before finally demonstrating a number of power peaks within the range commonly associated with $p$-modes. It must be remembered that the A-type \edit1{ and K-type} stellar sources provided flat and relatively featureless spectra, with no spectral slopes visible in their corresponding PSDs. Hence, the A- \edit1{ and K-}type PSDs (see, e.g., the upper \edit1{and middle} panel of Figure~{\ref{PSD_NGTS}}) show no agreement with the synthetic PSD depicted in Figure~{\ref{PSD_SIM}}, and serves as a further indicator that there is no nanoflare activity present on our A- \edit1{and K-}type stellar samples.

In a consistent manner with how the M-type stellar PSDs were processed, each of the 6100 synthetic lightcurves were examined and their corresponding primary frequencies, spectral slopes, and dominant frequencies (following detrending by the computed spectral gradients) were calculated. In order to more readily display these sets of measured parameters, we display them in Figure~{\ref{Heat Map}} in a two-dimensional format as a function of the power-law index, $\alpha$, and the corresponding decay timescale, $\tau$. This is similar to the intensity fluctuation statistical measurements depicted in Figure~{\ref{Stat_Map}}, only Figure~{\ref{Heat Map}} now displays the corresponding parameters extracted from the analysis of the synthetic PSDs. 

Figure~{\ref{Heat Map}} documents interesting behavior of the key Fourier-based parameters as a function of the power-law index, $\alpha$, and the corresponding decay timescale, $\tau$. As the power-law index increases, the spectral slopes (upper-left panel of Fig.~{\ref{Heat Map}}) associated with the PSDs begin to flatten. This is likely a consequence of increased energy being spread across the entire frequency spectrum as a result of the larger power-law indices \citep{Jess2020}. Previous work on turbulent cascades have revealed spectral slopes within the range of $-2 \leq \beta \leq -1$ in both solar and stellar plasma \citep{Podesta:2011, Huang:2017}, believed to be a feature of wave behavior. The spectral slopes found in our simulation outputs varies largely within this range ($-1.85 \leq \beta \leq 0.00$), but is a result of pure nanoflare signals, with no presence of strictly wave-based signatures. An explanation could be that nanoflares are individually low-energy events, but they occur very frequently all over the surface of a star. They may come together with an additive effect to form {(quasi-~)}periodic signals, as opposed to the breaking effect of a wave cascade. This cascade-like signal has been documented previously by \citet{Hudson:1991}, wherein solar nanoflare simulations produced a similar power spectrum cascade, but here we present the first evidence in stellar-specific simulations. This cascade signal is also similar to the `inverse magnetic cascade' process discussed in \cite{Christensson:2001}, who found a reverse turbulence effect in 3-D MHD simulations, lending support to an inverse cascade signal generated by magnetic behaviour. 

Furthermore, the primary frequency (lower-left panel of Figure~{\ref{Heat Map}}) is sensitive to the nanoflare decay timescale, rising from $\sim$0.7~mHz at the longest $e$-folding times ($\approx$500~s) to $\sim$1.4~mHz at the most rapid decay timescales ($\approx$10~s). Interestingly, once the PSDs have been detrended by their corresponding spectral slopes, the dominant frequencies (upper-right panel of Figure~{\ref{Heat Map}}) present are within the range of $2.7 - 4.2$~mHz. This frequency range is often synonymous with the presence of $p$-mode waves \citep{Andrews:1989, Andrews:1990a, Andrews:1990b}, even though our simulations contain no strict wave activity.

An interesting metric to benchmark how significant the power peaks are within the range of $1-5$~mHz involves the calculation of the percentage of the nanoflare spectral power equal to, or greater, than the corresponding power found in the synthetic noise PSD (lower-right panel of Figure~{\ref{Heat Map}}). We find that the spectral power arising from strictly nanoflare signatures is $10-100$\% greater than the corresponding (flat) noise power arising from a Poisson-based shot noise distribution. This can be seen in the upper panel of Figure~{\ref{PSD_SIM}}, whereby the power arising from nanoflare signals is above that corresponding to the noise floor.

Comparing the simulated PSD features to the M-type stars (see Table~{\ref{PSD_NGTS}}), we find overlaps with the two-dimensional maps shown in Figure~{\ref{Heat Map}} for a power-law index $\alpha  = 3.3 \pm  0.2 $ and a nanoflare decay timescale $\tau = 200 \pm  100$~s. These values are consistent with the first set ($\alpha=3.25\pm0.15$ and $\tau=200\pm100$~s) of plasma conditions extracted from the intensity fluctuation statistical distributions. Importantly, we do not find self-similar PSD results substantiating the second set ($\alpha=2.00\pm0.15$ and $\tau=200\pm100$~s) of plasma conditions extracted from the intensity fluctuation statistical distributions. This demonstrates the usefulness of employing both statistical and Fourier-based benchmarking of the observational and synthetic time series, since it has allowed us to alleviate a potentially ambiguous result found using just a single analysis method.

\subsection{Observed Stellar Nanoflaring Parameters }

Combining both the statistical and PSD benchmarks, we find evidence for stellar nanoflare activity across the sampled M-type stars for a power-law index $\alpha  = 3.25 \pm  0.20 $ and a decay timescale $\tau = 200 \pm  100$~s. Considering we are integrating over an entire stellar disk, we would expect some variation in the local plasma conditions, hence the relatively large uncertainties placed on the decay timescales. While the $e$-folding timescale is comparable to those put forward in solar studies, the power-law index is much higher than the range ($1.82 \leq \alpha \leq 1.90$) observed in solar plasma by \citet{jess:2019}. It also exceeds the full range ($1.35 \leq \alpha \leq 2.90$) reported across the literature for all solar events \citep{Berghmans:1998, Krucker:1998, Aschwanden:1999a, Parnell:2000, Benz:2002, Winebarger:2002, Aschwanden:2012, Aschwanden:2014, Aschwanden:2015}. 
\edit1{This also exceeds the typical range of stellar flare power-law indices $1.5 \le \alpha \le 2.7$ reported by, e.g., \citet{Hudson:1991, Robinson:1995, Robinson:1999, Kashyap:2002, Gudel:2003, gudel:2004, Welsh:2006, Reale:2016}.
 }

M-type stars are nearly or fully convective, \edit1{with more powerful magnetic activity than the Sun, leading to increased flare activity}. However, this alone cannot explain a higher power-law index, as a general boost to activity levels would enhance all frequencies and energies, thus preserving the same power-law index. Instead, it is possible that small-scale nanoflare energies in the range  $10^{22} - 10^{25}$~ergs are boosted disproportionately in these flare active stars.
\edit1{While low energy flares are likely to be governed by the same underlying physical processes \citep{Lu:1991}, and the power-law relationship is scale-free \citep[applying to both small and large flares;][]{Aschwanden:2019}, \citet{Robinson:1995} and \cite{Vlahos:1995} suggested that a discontinuity in the power-law indices of high and low-energy flare events would be an inherent feature of the self-organized criticality model of flaring (wherein small magnetic reconnections occur very frequently, each with the potential to setoff another reconnection nearby, causing an avalanche effect, and following a power-law distribution of energies). They suggest that while high energy flaring would exist at power-law indices of $\alpha  = 1.8$, the power-law index of low energy (i.e. micro and nanoflares) would range around $3 \leq \alpha \leq 4$. This is in agreement with our stellar nanoflaring power-law index of $\alpha  = 3.25 \pm  0.20 $}. Another explanation for this enhanced rate of small-scale flare activity in M-dwarf stars could lie in the reconnection process itself. \citet{Tsuneta:2004} suggested that low energy  (pico-/nano-)flares may occur more favorably via Sweet-Parker reconnection (instead of Petschek processes). If such flare stars have lower Lundquist numbers (i.e., higher plasma resistivity) with respect to the Sun, then this may help explain the enhanced nanoflare rates found in our present study. The mostly convective atmosphere of these flare stars may be able to modify the underlying Lundquist number, allowing for enhanced low-energy nanoflare rates via Sweet-Parker reconnection, but not modifying the rates of the higher energy events that will proceed (as normal) via Petschek reconnection processes. This enhanced nanoflaring may also be linked to the dynamo in these stars. \edit1{The M-stars in this study sit on the boundary of fully convective atmospheres. While the spectral sub-type where full convection begins is still under debate, estimates are in the range M3 and above \citep{Wright:2016} to more recent studies suggesting M2.1 to M2.3  \citep{Mullan:2020}}. Fully convective stars lack the tachocline between convective and radiative zones which powers the solar dynamo . A dynamo powered by helical turbulence is believed to operate in these fully convective stars \citep{Durney:1993, Browning:2008, Pipin:2009}. This may operate in tandem with the enhanced Sweet-Parker reconnection, through altering the Lundquist number. Investigating the power-law indices of nanoflaring signatures for stars either side of this convective boundary (i.e., M1 and M5) would allow us to test this theory in a future study. 

\section{Conclusions}

We have employed a combination of statistical and Fourier-based analysis techniques to search for evidence of nanoflare activity in M-type stars observed by NGTS. The intensity fluctuation distributions of the M-type stars revealed both negative median offsets and positive Fisher skewness values, highlighting the presence of impulsive intensity rises, followed by exponential decays, trapped within the noise envelop of their corresponding lightcurves. To validate these signatures, we examined complementary \edit1 {A-type non-flare active and K-type low-activity }stars, which demonstrated zero median offsets, alongside very minor Fisher skewness values, highlighting the more symmetric composition of these A- \edit1{and K-} type time series that are devoid of nanoflare signatures.

Previous studies have observed periodic phenomena in M-type stars that has been interpreted as evidence of $p$-mode wave activity. To investigate whether nanoflare signatures, which are governed by a power-law index, may contribute to similar (quasi-)periodicities we calculated power spectral densities (PSDs) of the NGTS time series. Long-duration successively acquired time series (2316 individual datapoints) were employed to maximize the frequency resolution. We found contrasting spectral features between the A- \edit1{ K-} and M-type time series. The A-type spectra had flat power trends representative of pure shot noise distributions, \edit1{The K-type PSD was flat apart from frequency enhancements across the range of $ \approx 1-10$~mHz, indicative of $p$-mode wave signatures, as we would expect from a Solar-like K-type star }. By contrast, the M-type spectra revealed spectral slopes and frequency enhancements across the range of $ \approx 1-10$~mHz. As a result, it was unclear whether the frequency enhancements were a result of nanoflare activity governed by a power-law relationship, or the capture of $p$-mode wave signatures. To investigate this further, we employed Monte Carlo models of nanoflare activity to examine whether pure flare signatures have the ability to manifest as spectral power enhancements in their corresponding PSDs.

A grid of 6100 Monte Carlo models were constructed that replicate the exposure time, cadence, and duration of the NGTS observations, but with each resulting time series generated from different combinations of power-law indices, $\alpha$, and flare decay timescales, $\tau$. Each of the time series were added to synthetic shot noise distributions to simulate realistic NGTS lightcurves. These were examined using identical statistical and Fourier-based techniques, with the results cross-correlated to the observational findings. Importantly, we found evidence that time series comprised of nothing but impulsive nanoflare signatures and Poisson-based shot noise are able to demonstrate spectral power peaks across the frequency range $ \approx 1-10$~mHz, suggesting that previously detected $p$-mode signatures may actually arise from nanoflare activity in the host star. Combining both the statistical and PSD benchmarks, we find evidence for stellar nanoflare activity across the sampled M-type stars for a power-law index $\alpha  = 3.25 \pm  0.20 $ and a decay timescale $\tau = 200 \pm  100$~s.

Looking to the future, higher cadence observations from instruments such as HiPERCAM \citep{Dhillon2016} may allow for the more rapid accumulation of suitable number statistics, plus the ability to investigate potential nanoflare signals across a number of different color photometry bands.\edit1{ As the nanoflaring parameters observed in the Sun by \citet{jess:2019} varied with wavelength, we would expect a similar result for stellar nanoflares. Such multi-wavelength observations could allow for a limited analysis of how the nanoflare signals differ throughout the stellar atmosphere. Investigating young Sun-like stars with HiPERCAM is also an area of interest, as the high cadence and multi-color observation can overcome the difficulties of increased flare contrast on these young and active stars. Due to their highly active x-ray emission and coronal temperatures  \citep{Johnstone:2015}, we expect a very high degree of nanoflare activity, possibly leading to stellar coronal heating via nanoflaring.} Furthermore, a follow up observational campaign could leverage the large sky sampling of the NGTS to examine the presence of nanoflare signatures on other spectral classifications, particularly M1 and M5 spectral types, and investigate how the convective boundary affects the nanoflare power-law indices. \edit1{This larger star sample could also investigate the source of the dip below the idealized Gaussian at approximately  $-0.90~\sigma_{N}$ in the statistical distribution of the M-stars.} To further improve the Fourier-based PSD analyses, we propose a more continuous observational platform that will further increase the frequency resolution possible, e.g. the Transiting Exoplanet Survey Satellite \citep[TESS;][]{Ricker:2014}, which can operate in both 240~s and 20~s cadences. The 20~s cadence data is part of the extended mission program that will begin operations in July 2020. The obvious advantages of space-based observations would allow us to minimize any high frequency (scintillation) noise present in the stellar lightcurves, while also allowing for a much higher frequency resolution in the subsequent PSD analyses. Longer-duration observations have been proposed to study stellar oscillations in greater detail \citep{Ball:2018}, and this capability would extend the same advantages to our nanoflare PSD analyses.

\acknowledgments
\noindent C.J.D. and D.B.J. wish to thank Invest NI and Randox Laboratories Ltd. for the award of a Research and Development Grant (059RDEN-1) that allowed the computational techniques employed to be developed. 
D.B.J. would like to thank the UK Science and Technology Facilities Council (STFC) for an Ernest Rutherford Fellowship (ST/K004220/1), in addition to a dedicated standard grant (ST/L002744/1) that allowed this project to be undertaken. S.J.C would like to thank the UK Science and Technology Facilities Council (STFC) for an Ernest Rutherford Fellowship (ST/R003726/1)
This project is based on data collected under the NGTS project at the ESO La Silla Paranal Observatory. 
The NGTS facility is operated by the consortium institutes with support from the UK STFC under projects ST/M001962/1 and ST/S002642/1. 
P.J.W., D.R.A., and R.G.W. acknowledge support from STFC consolidated grants ST/L000733/1 and ST/P000495/1.
D.B.J. wishes to acknowledge scientific discussions with the Waves in the Lower Solar Atmosphere (WaLSA; \href{https://www.WaLSA.team}{www.WaLSA.team}) team, which is supported by the Research Council of Norway (project no. 262622) and the Royal Society (award no. Hooke18b/SCTM). This research has made use of data obtained from the 4XMM XMM-Newton Serendipitous Source Catalog compiled by the 10 institutes of the XMM-Newton Survey Science Centre selected by ESA
%

\facilities{Next Generation Transit Survey (NGTS)}




\clearpage

\appendix

\setcounter{table}{3}
\renewcommand{\thetable}{A\arabic{table}}
\section{Stellar Parameters}
\label{sec:Appendix_A}

Additional stellar parameters, including the RA and Dec for each star are described in Table~{\ref{tab:StellarParam}}.

\begin{sidewaystable*}[h]

\centering
\resizebox{\linewidth }{!}{%

\begin{tabular}{@{}l|ccccccccc@{}}

\toprule
SP Type & M2.5v & M3V & M2.5V & A5V & A5V & A7V  & K2V & K2V & K2V\\ 
NGTS ID & NGTS J030047.1-113651 &NGTS J030415.6-103712 & NGTS J031800.1-212036 & NGTS J025840.5-120246 & NGTS J030958.4-103419& NGTS J030129.4-110318 & NGTS J030000.7-105633 &NGTS J030848.9-112217 & NGTS J030538.9-114145 \\
GAIA ID & 5160579407177989760 & 5160771340676667776 & 5099679725858611840 & 5160183681775577472 & 5165979280580778624 & 5160773569763964416 & 5160700765773865600 &  5165722991292368384 &  5159884962505997184 \\
TIC ID & 141307298 & 23138344 & 92249704 & 98757710 & 23221987 & 141309114 & 141287385 & 23192572 & 23169095\\
RA & $45.196372\degree $ & $ 46.065155\degree $ & $ 49.500502\degree  $ & $ 44.66885\degree $ & $ 47.493582\degree $ & $ 45.372827\degree  $  & $45.003119\degree $ & $ 47.203979\degree $ & $ 46.412072\degree$  \\
Dec & $ -11.614197\degree $ & $  -10.620268\degree $ & $ -21.343482\degree $ & $  -12.046304\degree $ & $  -10.572118\degree $ & $ -11.055091\degree $ & $ -10.942633\degree $ & $  -11.371446\degree $ & $ -11.696004\degree $  \\
Mass ($M_\odot$) & $ 0.40 \pm 0.02 $ & $ 0.55 \pm 0.02 $ & $ 0.40 \pm, 0.02$ & $ 2.28 $ & $ 1.27 \pm 0.21 $ & $ 1.81  \pm 0.29 $  & $ 0.77 $  & $ 0.78 $ & $ 0.78 $\\
Radius ($R_\odot$) & $ 0.41 \pm 0.01$ & $ 0.55 \pm  0.02$ & $ 0.41 \pm  0.01 $ & $ 3.03 $ & $ 1.16 \pm  0.05 $ & $ 1.65 \pm  0.09 $ & $ 0.85 $  & $ 1.16 $ & $ 0.78$ \\
Luminosity ($L_\odot$) & $  0.021 \pm  0.005 $ & $  0.036 \pm  0.009$ & $ 0.021\pm  0.005 $ & $54.916  $ & $ 2.0 \pm  0.1 $ & $  8.6\pm  0.8 $ & $ 0.342 $  & $ 0.647 $ & $ 0.356$  \\ 
Distance (pc) & $  67.5 \pm  0.4 $ & $  125.6 \pm  1.6 $ & $ 59.6\pm  0.2 $ & $ 3464.1\pm 477.9 $ & $ 690.6 \pm  11.8 $ & $  737.7 \pm  30.5 $ & $  375.1 \pm  10.3 $ & $  532.2 \pm  12.9 $ & $  381.1 \pm  5.6 $ \\ 
Macroscopic Flare Rate (Flares per Hour) & $ 0.012 $  &$ 0.027 $ & $ 0.003 $ &$ 0 $  &$ 0 $ &$ 0 $ &$ 0 $ &$ 0 $ &$ 0 $ \\
$ \log\left(\frac{L_x}{L_{Bol}}\right ) $ & $ -3.09 \pm 0.21$ & &  &  & &  & & & \\

\bottomrule

\end{tabular}}

\caption{ The Spectral type, NGTS identifier, Gaia source ID, Tess Input Catalog (TIC) ID, RA, Dec, Stellar Mass (in Solar mass units), Stellar Radius (in Solar radi units), Stellar Luminosity (in Solar luminosity units), Distance (in parsecs), Macroscopic Flare Rate (per hour) and the ratio $ \log\left(\frac{L_x}{L_{Bol}}\right) $ for the stars used in the analysis.  The Stellar masses, radi. and luminosity data is from the Tess Input Catalog release  V8.  \citep{Stassun:2018}  \edit1{ The ratio $ \log\left(\frac{L_x}{L_{Bol}}\right) $ is calculated by comparing the log ratio of the x-ray luminosity as observed by 4XMM XMM-Newton Serendipitous Source Catalog \citep{Webb:2020}, with the stars luminosity. A ratio of  $\log\left(\frac{L_x}{L_{Bol}}\right)  \sim -3 $ is expected for x-ray saturated, young and active M stars \citep{Kastner:2003} . }  }
\label{tab:StellarParam}

\end{sidewaystable*}

\clearpage

\section{Power Spectral Density Compared}
\label{sec:Appendix_b}
\setcounter{figure}{7}
\renewcommand{\thefigure}{B\arabic{figure}}

\begin{figure}[H]
  \centering
  \includegraphics[ clip=true, width=\textwidth, angle=0]{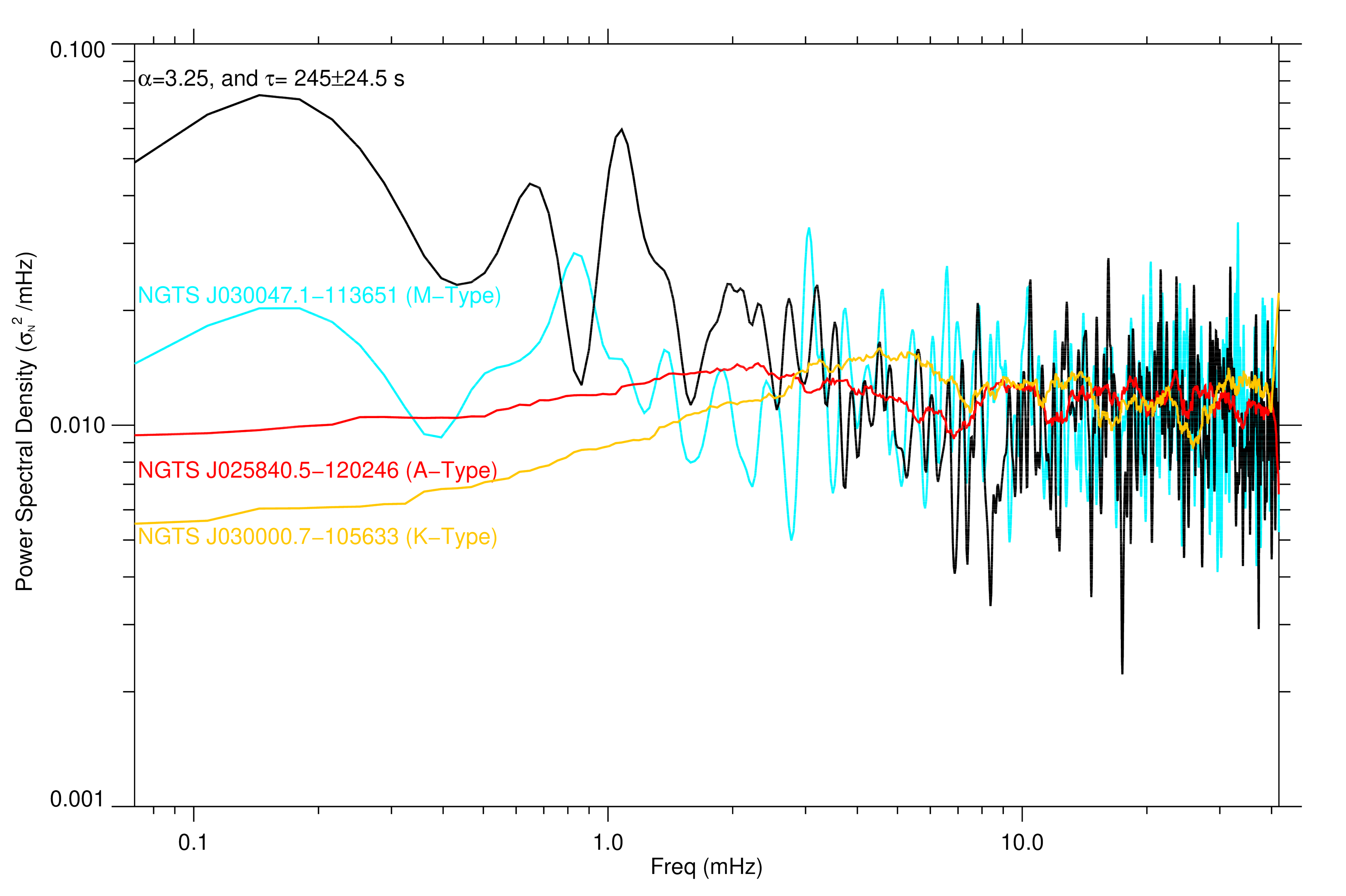}
      \caption{\edit1{The Fourier power spectral density (PSD) trendlines calculated over $\pm$6 frequency elements ($\pm$0.427~mHz) for example A- \edit1{K-} and M-type stellar sources NGTS J025840.5-120246 (red line), \edit1{NGTS J030000.7-105633  (orange line)} and NGTS J030047.1-113651 (blue line), and a modeled time series corresponding to a power-law index $\alpha=3.25$ and a flare decay timescale $\tau= 245 \pm 24.5$~s (black line), displayed in normalized units of $\sigma_{N}^{2}/\text{mHz}$. It can be seen that the A-type \edit1{and K-type} spectra are relatively flat across all frequencies with no evidence of distinct peak frequencies. \edit1{The K-type does show some slight power enhancement between $\approx 1-10$~mHz, consistent with stellar $p$-mode oscillations, as have been previously observed in K-type Solar-like stars \citep[e.g.][]{Chaplin:2009}  }. Contrarily, the M-type PSD has a primary power peak at $\approx$0.8~mHz, followed by decreasing spectral power exhibiting a spectral slope of $\beta = -0.30 \pm 0.05$, followed by numerous power peaks in the range of $3-10$~mHz, consistent with previous links to stellar $p$-mode spectra. The synthetic PSD is remarkably similar to the NGTS M-type stellar source, with peaks and spectral slopes in the same range and magnitude (see Figure~{\ref{Heat Map} for the full range of peak frequencies and spectral slopes in modeled time series PSDs.}}  }
         \label{PSD_ALL}
  \end{figure}

\edit1{Figure~{\ref{PSD_ALL}} shows the trendlines (calculated over $\pm$6 frequency elements or $\pm$0.427~mHz) of  Fourier power spectral densities (PSDs) for the example A \edit1{K, }and M stars, as well as for a modeled time series with a power-law index $\alpha=3.25$ and a flare decay timescale $\tau= 245 \pm 24.5$~s. This plot highlights the agreement in the observational M-type and modeled time series PSDs, with comparable spectral slopes of approximately $\beta = -0.30 \pm 0.05$, and peaks around  $\approx$0.8~mHz . This is in contrast with the A- \edit1{and K-}type PSD, which are relatively flat and featureless by comparison.}

\section{Observational Considerations}
\label{sec:Appendix_C}
\edit1{The detectability of nanoflare signals via statistical and periodic analyses are dependent on the underlying observational parameters, including,}
\begin{itemize}
\edit1{  \item Time Series Length: The statistical analysis is dependent on the number of frames, $N$. The error in statistical analyses scales with $\sqrt{N} $, while the signal scales with $N$. Periodic analysis is also dependent on the time series length, but crucially on the length of successive uninterrupted frames. Increasing the duration of the observations will provide increased frequency resolution. The periodic signal also benefits from increased number statistics, as the number of nanoflares captured increases with longer observing sequences, hence providing more accurate quantification of any associated periodicities. We have investigated  modeled lightcurves (which are not subject to day/night cycles), and found increasing the number of successive frames had the effect of increasing the ratio of nanoflare power above the noise floor  in the range $1-5$~mHz (i.e., the lower panel of Figure 7). As a result, the nanoflare periodic signatures became more prominent over the noise. We expect space-based (e.g., TESS) observation to allow us to uncover more of the underlying spectral slopes, particularly for the highest power-law values.
\item Cadence: Shorter cadences will allow for increased Nyquist frequencies to better resolve rapid and short-lived periodic signatures. Sub-second cadences \citep[e.g., HiPERCAM, with exposures on the order of milliseconds;][]{Dhillon2016} could allow for a very large frequency range and excellent number statistics to be achieved in a very short observation window. 
\item Apparent Magnitude: As the observed magnitude increases, the scintillation noise begins to increase also (this is not an issue with space-based observations). This would affect the frequency distribution of the noise, since the scintillation introduces a frequency-dependent noise component that needs to be considered. Searching for nanoflares embedded within this more complex noise distribution would require the seeding of a scintillation model into the numerical simulations. A future study could explore high-magnitude stars, to determine whether the increased scintillation is balanced by the increased nanoflare signal, or future space observations (e.g., TESS) could mitigate this entirely. However, the star itself should still be of low intrinsic stellar brightness; see below.
\item Intrinsic Stellar Brightness: Brighter stars have increased quiescent flux, and therefore a more pronounced noise floor that must be combated when searching for nanoflare signals on top of this brighter background. This means the contrast between the nanoflare signals and the background becomes a challenging issue. Even at solar-like luminosities, the detection of microflare energies becomes difficult, let alone nanoflares on stellar sources that cannot be spatially resolved.
}

\end{itemize}

  
\bibliography{bib.bib}{}
\bibliographystyle{aasjournal.bst}



\end{document}